\documentclass[prx,amsmath,amssymb,notitlepage,longbibliography,eqsecnum]{revtex4-2} 

\pdfoutput=1

\usepackage{graphicx,bm}

\usepackage[tight]{subfigure} 

\usepackage{color} 

\usepackage[papersize={8.5in,11in}]{geometry}
        
\usepackage{color}
\definecolor{darkblue}{rgb}{0.,0.,0.4}
\definecolor{darkred}{rgb}{0.5,0.,0.}
\definecolor{BlueViolet}{RGB}{138,43,226}
\definecolor{SkyBlue}{RGB}{30,144,255}
\definecolor{DarkGreen}{RGB}{0,100,0}
\usepackage[pdftex,colorlinks=true,linkcolor=darkblue,citecolor=blue,urlcolor=darkred]{hyperref}

\usepackage{amsfonts,bbold}
\usepackage{upgreek}
\usepackage{slashed}
\usepackage{latexsym}
\usepackage{ulem}

\usepackage{wasysym}			
\usepackage{amsthm}
\usepackage{xcolor}
\usepackage{makecell,hhline}

\newcommand{\nn}{\nonumber \\}
\newcommand{\be}{\begin{equation}}
\newcommand{\ee}{\end{equation}}

\renewcommand{\vec}[1]{{\bf #1}}

\newcommand{\Tr}{{\rm Tr}\,}

\newcommand{\rmi}{{\rm i\,}}

\renewcommand{\epsilon}{\varepsilon}

\begin{document}
\title{Fate of superconductivity in three-dimensional disordered Luttinger semimetals}

\author{Ipsita Mandal}
\affiliation{Max-Planck Institute for the Physics of Complex Systems\\
Noethnitzer Str.~38, 01187, Dresden, Germany}

\begin{abstract}
%%%%%%%%%%%%%%%%%%%%%%%%
Superconducting instability can occur in three-dimensional quadratic band crossing semimetals only at a finite coupling strength, due to the vanishing of density of states at the quadratic band touching point. Since realistic materials are always disordered to some extent, we study the effect of short-ranged-correlated disorder on this superconducting quantum critical point, using a controlled loop-expansion applying dimensional regularization. The renormalization group (RG) scheme allows us to determine the RG flows of the various interaction strengths, and shows that disorder destroys the superconducting quantum critical point. In fact, the system exhibits a runaway flow to strong disorder.
\end{abstract}

\maketitle
\tableofcontents

%==========================================================================

\section{Introduction}

Three-dimensional isotropic systems with a quadratic band touching (QBT) point, supplemented by Coulomb interactions, were studied by Abrikosov back in 1971 with the renormalization group (RG) technique in $4-\epsilon$ spatial dimensions \cite{Abrikosov}. It was argued that the long-ranged Coulomb interactions may stabilize a {\textit{non-Fermi liquid}}  ground state \cite{Abrikosov,MoonXuKimBalents}. Such a system is then possibly the simplest example of a non-Fermi liquid.
Recently, there has been a revived interest in these systems~\cite{MoonXuKimBalents,Herbut,Herbut2,Herbut3,LABIrridate} motivated by their relevance to pyrochlore iridates $\text{A}_ 2\text{Ir}_ 2\text{O}_ 7$ , where A is a lanthanide element \cite{pyro1,pyro2}. Furthermore, if the spin-orbit coupling is strong enough in three-dimensional gapless semiconductors, then it can cause the Fermi level to lie at a QBT point \cite{Beneslavski}, and such a model is indeed relevant for gray tin (HgTe). These systems have been dubbed as ``Luttinger semimetals'' \cite{igor16}, since the low-energy electronic degrees of freedom are captured by the Luttinger Hamiltonian  of inverted
band gap semiconductors \cite{luttinger,murakami}. The interplay of disorder and Coulomb interactions at the QBT has been investigated in Ref.~\cite{LaiRoyGoswami,rahul-sid,ipsita-rahul,*ips-rahul-errata}, where the RG flows of the coupling strengths show that disorder is a relevant perturbation to Abrikosov's non-Fermi liquid fixed point, and that the disordered problem undergoes a runaway flow to strong disorder \cite{rahul-sid,ipsita-rahul,*ips-rahul-errata}.

When the chemical potential is at the QBT point, an attractive four-fermion interaction can lead to a superconducting instability only at a finite coupling strength due to the vanishing density of states at QBT, leading to the possibility of a quantum critical point. Such a scenario for a clean system and in the absence of Coulomb interactions has been studied in Ref.~\cite{igor16} and a stable quantum critical point for s-wave superconductivity identified. Neglecting the Coulomb interaction is justified if it is rendered sufficiently weak by a large dielectric constant of the material. In this work, we examine the fate of this superconducting quantum critical point in the presence of disorder. It is worth mentioning that in the context of two-dimensional systems, interplay of superconducting critical points and disorder has been previously studied in Ref.~\cite{rahul1,rahul2,2dsc1} for the case of massless spinful Dirac fermions (relevant for graphene).  
The half-Heusler compound YPtBi is a noncentrosymmetric multiband superconductor with QBT point and a promising candidate for hosting topologically nontrivial superconducting states in three dimensions. The bulk and surface states of two prototypical pairing states in YPtBi, one preserving time-reversal symmetry, the other breaking it, have been studied in Ref.~\cite{carsten1}. In Ref.~\cite{carsten2}, the authors have showed that for a centrosymmetric superconductor with a QBT point and a broken time-reversal symmetry, the low-energy excitation spectrum has two-dimensional Bogoliubov Fermi surfaces in the bulk instead of point or line nodes. 
Furthermore, instabilities of various non-Fermi liquid scenarios \cite{nfl0,nfl1,nfl2,sur} towards superconductivity have been studied extensively in the literature \cite{ips2,ips3,Max,ips-sc-isn,ipsc2}.

This paper is structured as follows: In Sec.~\ref{model0}, we introduce the basic non-interacting model and add the superconducting s-wave pairing channel. In Sec.~\ref{clean}, we revisit the existence of the infrared stable superconducting quantum critical point using the minimal subtraction scheme of RG. In Sec.~\ref{disorder}, we study the interplay of superconductivity and disorder by the same RG scheme. We show that the superconducting quantum critical point is destroyed, and the problem continues to flow to strong disorder. We conclude with some discussion and overview in Sec.~\ref{discussion}. The appendices contain technical results used in the computations.

%%%%%%%%%%%%%%%%%%%%%%%%%%%%%%
\section{Model}
\label{model0}

We consider a model for three-dimensional quadratic band crossings, where the low energy bands form a four-dimensional representation of the lattice symmetry group \cite{MoonXuKimBalents}. Then the standard $\vec{k} \cdot \vec{p}$ Hamiltonian for the non-interacting system, in the absence of disorder, can be written by using the five $4\times 4$ Euclidean Dirac matrices $\Gamma_a$ as \cite{Herbut}:
 \begin{equation}
 \mathcal{H}_0 = \sum_{a=1}^5 d_a(\vec{k}) \,  \,\Gamma_a   +\xi\,  {k^2}  \,,
\label{bare}
 \end{equation}
%%%%%%%%%%%%%%%%
with the $\,\Gamma_a$ providing one of the (two possible) irreducible, four-dimensional Hermitian representations of the five-component Clifford algebra defined by the anticommutator $\{ \,\Gamma_a, \,\Gamma_b \} = 2\, \delta_{ab}$. In $d=3$, the space of $4\times 4$ Hermitian matrices is spanned by the identity matrix, the five $4\times 4$ Gamma matrices $\Gamma_a$ and the ten distinct matrices $\Gamma_{ab} = \frac{1}{2\,i}\, [\Gamma_a, \Gamma_b]$. The five anticommutating gamma-matrices can always be chosen such that three are real and two are imaginary \cite{igor12}. We choose a representation in which $(\Gamma_1, \Gamma_2, \Gamma_3)$ are real and $(\Gamma_4, \Gamma_5 ) $ are imaginary.
The five functions $  d_a(\vec{k})$ are the real $\ell=2$ spherical harmonics, with the following structure:
 \begin{eqnarray}
\label{ddef}
  d_1(\vec{k}) &=& \sqrt{3}\, k_y \,k_z\,,
\quad   d_2(\vec{k}) =  \sqrt{3}\, k_x\, k_z\, ,\quad
 d_3(\vec{k}) =  \sqrt{3} \,k_x\, k_y\, ,
\nonumber\\
  d_4(\vec{k})&=&\frac{\sqrt{3}  \,  (k_x^2 - k_y^2) }{2}\,, 
\quad   d_5(\vec{k}) = \frac{2\, k_z^2 - k_x^2 - k_y^2}{2} \,.
\end{eqnarray}
The isotropic $\xi \, k^2 $ term with no spinor structure introduces band-mass asymmetry to the band-structure.

%%%%%%%%%%%%%%%%%%%%%%%%%%%%%%%%%%%%%%%%
\subsection{Interactions for generating superconductivity}

In this subsection, we review the derivation of the effective action which can lead to a superconducting instability, as discussed in Ref.~\cite{igor16}.
In order to generate Cooper pairing, we add local attractive interactions, such  that the zero-temperature Euclidean action is given by:
\begin{align}
 \label{qbt5} \mathcal{S}[\psi] = \int \mbox{d}\tau \,\mbox{d}^dx \Bigl [ \psi^\dagger\left  (\partial_\tau +  \mathcal{H}_0 \right )\psi + V (\psi^\dagger \psi)^2 \Bigr ],
\end{align}
where $\tau$ is the imaginary time, $d$ is the number of spatial dimensions, $V <0$ is an attractive coupling constant, and $\psi(\tau,\vec{x})=(\psi_1,\psi_2,\psi_3,\psi_4)^{\rm T}$ is the four-component Grassmann field.
The engineering dimension of $V$ is $[V] = 2 - d$, which means that it is an irrelevant coupling in $d=3$. As a result, the conventional BCS pairing for infinitesimally small attractive $V$ is not possible. Physically, this is because the density of states vanishes at the QBT \cite{nozieres}. Nevertheless, there is a possibility of quantum phase transition at a sufficiently large value of coupling, where the system may lower its ground state energy by opening a gap at the Fermi level.

It was shown in Ref.~\cite{igor16} that there can be two competing superconducting orders: $\phi = \langle \psi^{\rm T}\, \,\Gamma_{45} \,\psi \rangle$ and $\tilde \phi = \langle \psi^{\rm T}\, \,\Gamma_{45} \,\,\Gamma_a\, \psi \rangle$, corresponding to s-wave and d-wave components, respectively. 
 It was also argued that s-wave ordering is energetically preferred as the gap is rotationally symmetric. Therefore, we set $\tilde \phi=0$ and investigate only the case of s-wave superconducting instability.

To capture the above physics, we can write the effective Lagrangian as:
 \begin{align}
 \label{qft1} 
    \mathcal L (\psi,\phi) = & \,\psi^\dagger
    \left [\partial_\tau +   \vec d\left (-\rmi \nabla \right )\cdot \vec \Gamma 
   -  {\xi} \, \nabla^2 \right ]\psi 
 + \phi^*\left (y\,\partial_\tau-c^2  \, \partial_\tau^2-\nabla^2 +r \right )\phi +\zeta \,|\phi|^4 \nn
 &
 + g \left ( \phi\, \psi^\dagger \,\Gamma_{45}\, \psi^* + \phi^* \psi^{\rm T} \,\Gamma_{45} \,\psi \right) .
\end{align}
The tuning parameter $r$, as usual, is proportional to $\left (V - V_{\rm c} \right )$, where $V_{\rm c} < 0$ is the critical value of the attractive interaction.
Hence, the quantum critical is located at $r=0$. The complex bosonic field $\phi$ is coupled to the fermions as a Majorana mass \cite{herbut2013}. The fields and the time coordinate have been rescaled such that the coefficients of the terms $\psi^\dagger \, \partial_\tau \psi$, $\left[ \psi^\dagger\, \vec d\left (-\rmi \nabla \right )\cdot \vec \Gamma  \, \psi \right ]$, and $ \left [- \phi^* \, \nabla^2 \, \phi \right ] $ are unity. We set $r=0$, assuming the theory to be close to its critical point.

\subsection{Engineering dimensions}

Let us determine the engineering dimensions of all the fields and coupling constants at the non-interacting Gaussian fixed point ($g=\zeta=0$) from the kinetic term with $[x]=-1$. Then, from the fermion dispersion, we get $[\tau]=-2$, leading to $[\psi(x)] =[\phi(x)]= \frac{d}{2}$ and $[\psi(P)]=[\phi(P)]=-\frac{d+4}{2}$. Finally, $[\xi] =[y]=0$, $[c]=- 1$, $[g]=\frac{4-d}{2}$ and $[\zeta] = 2-d $. Hence, for $d=4$, the coupling $g$ is marginal.
Since $c$ and $\zeta$ are irrelevant for any dimension $d > 2$, we drop them.

Therefore, we study the s-wave superconducting quantum critical point of the system by generalizing the theory to $d=4-\epsilon$ spatial dimensions (assuming $0 < \epsilon \ll 1$) in terms of the critical ($r=0$) effective action
\begin{align}
 \label{qft7}
 \mathcal  S_{0} & =    \int d\tau\, d^d x \,
\Big [  {\psi}^\dagger
\left \lbrace  \partial_{ {\tau}}+ d_a\left (-\rmi \nabla \right )\,\Gamma_a- {\xi} \, \nabla^2  \right \rbrace  {\psi} 
  + {\phi}^*\left ( {y}\,\partial_{ {\tau}}- \nabla^2 \right ) {\phi}\nn
& \hspace{ 2.3 cm}
+  {g}\,\mu^{\epsilon/2} \left  ( {\phi} \,{\psi}^\dagger \,\Gamma_{45}\, {\psi}^* +  {\phi}^* \, {\psi}^{\rm T} \,\Gamma_{45}\,  {\psi}\right )
\Big ]  \,, \nn
\end{align}
which includes all the relevant and marginal couplings at the Gaussian fixed point. A mass scale
$\mu$ is introduced to make $g$ dimensionless.

%%%%%%%%%%%%%%%%%%%%
\section{Self-energies and beta functions for the clean case}
\label{clean}

In this section, we compute the self-energies of the fermions and complex bosons, generated due to the interaction between them. We use these results in the minimal subtraction scheme to determine the beta functions for the RG flows. Although similar calculations have already been done in Ref.~\cite{igor16}, we find it necessary to rederive those because our RG scheme is different from Ref.~\cite{igor16}, and also because the numerical factors obtained differ from the previous calculation.

We will consider the RG flow generated by changing $\Lambda$, which is the ultraviolet cut-off for the spatial momenta, by requiring that low-energy observables are independent of it.  This is equivalent to a coarse-graining procedure of integrating out high-energy modes . We note that for quadratic dispersion, the ultraviolet cut-off for energy is $\sqrt \Lambda $. When the loop-diagrams have a divergent dependence on $\Lambda$, this turns into a pole in $\epsilon$ in the dimensional regularization scheme, where we perform the energy and momentum integrals by integrating from $-\infty$ to $\infty$ setting $d=4-\epsilon$. For the angular integrals, we use the ``Moon-scheme" described in Sec.~\ref{angular}.

\subsection{One-loop calculations}
%%%%%%%%%%%%%%%%%%%%%%%

 Let $\Sigma_1 $ and $\Pi_1$ denote the one-loop corrections to the fermion and boson self-energies, respectively, where we use the sign convention where the self-energy subtracts the bare action in the dressed
propagator as $G(P) =\frac{1}{G_0^{-1}(P) - \Sigma_1 (P) } $ and $D(P ) =\frac{1}{D_0^{-1}(P) - \Pi_1 (P) }   $.
Here we have used the convention $P \equiv(\vec{p}, p_0)$, and denoted the zeroth order boson and fermion propagators by:
\begin{align}
G_0(P) = \frac{1} {\left (   i\, p_0 + \xi\, p^2\right )\mathbb{1}_N +\vec d (\vec{p})\cdot \vec \Gamma}
= \frac{ -\left (   i\, p_0 + \xi\, p^2\right )\mathbb{1}_N +\vec d (\vec{p})\cdot \vec \Gamma}
{ p_0^2 +\left (1- \xi^2\right ) p^4 -2\,i\, p_0\, \xi\,p^2 } \,,
\label{fermion-prop}
\end{align}
and
\begin{align}
D_0(P) = \frac{1} {i\, y \, p_0 + p^2 } \,.
\label{boson-prop}
\end{align}
Note that we have generalized to $N$ fermion components (flavors), where $N$ is a multiple of four.
In the Feynman diagrams, we will represent the fermion and boson propagators by solid and dashed lines, respectively.

\begin{figure}
\centering
\subfigure[]{\includegraphics[width = 0.35 \columnwidth]{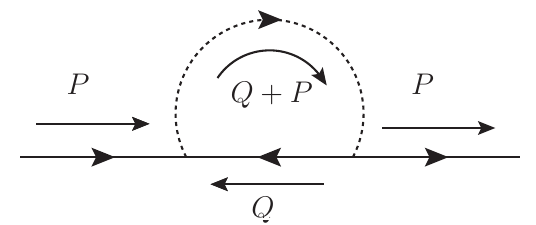}} \hspace{2 cm}
\subfigure[]{\includegraphics[width = 0.40\columnwidth]{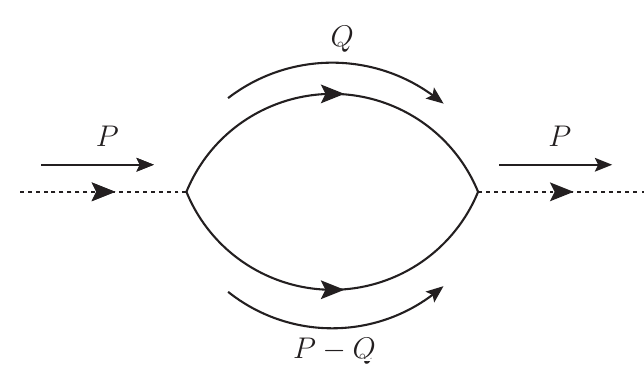}}
\caption{\label{self-en-fig} The one-loop (a) fermion and (b) boson self-energy diagrams.}
\end{figure}

Let us now compute one-loop self-energy corrections. 
A sample contraction of terms leading to the fermion self-energy can be written as:
\begin{align}
& \phi^* \left( \psi^T \,
\Gamma_{45} \, \psi \right) \phi \left( \psi^\dagger \,\Gamma_{45} \, \psi^* \right) 
%%%%%%%%%%%
= \left( \phi^* \phi\right)
 \psi_\alpha \,\left( \Gamma_{45} \right)^{\alpha,\beta}\,\psi_\beta\,
 \psi^*_\gamma \,\left( \Gamma_{45} \right)^{\gamma,\delta}\,\psi^*_\delta
%%%%%%%%%%%%%%%%%%%
 = -\psi^*_\delta \left[ \left( \Gamma_{45} \right)^{\delta, \gamma} 
\left( G_0^T \right)_{\gamma, \beta}
\left( \Gamma_{45} \right)^{\beta, \alpha } \, D_0
 \right ] \psi_\alpha \,.\nonumber
\end{align}
There are two such terms possible.
%%%%%%%%%%%%%%%%%%%%%%%%%%%%%%%%%%%%%%%%%%%%%%%%%%%%%5
Similarly, a sample contraction of terms leading to the boson self-energy can be written as:
\begin{align}
&\phi^* \left( \psi^T \,
\Gamma_{45} \, \psi \right) \phi \left( \psi^\dagger \,\Gamma_{45} \, \psi^* \right) 
 = \phi\,\phi^*
  \left[ \psi_\alpha \left( \Gamma_{45} \right)^{\alpha, \beta} \psi _\beta \right ]
  \left [ \psi^*_\gamma
\left( \Gamma_{45} \right)^{\gamma,\delta } \, \psi^*_{\delta} \right ] \nn
%%%%%%%%%%%%%
& =  \phi \,\phi^* \left[ 
\left( G_0^T \right)_{\delta,\alpha}
\left( \Gamma_{45} \right)^{\alpha, \beta} 
\left( G_0 \right)_{\beta,\gamma}
\left( \Gamma_{45} \right)^{\gamma,\delta } 
\right ]
%%%%%%%%%%
=\phi \,\phi^* \Tr \left[ 
 G_0^T \, \Gamma_{45} \, G_0 \,\Gamma_{45} \right ] .\nonumber
\end{align}
Again, there can be two such terms.

The explicit expressions for the fermion and self-energies are given by:
%%%%%%%%%%%%%%%%%%%%%%%%%
\begin{align}
 \label{rg7} \Sigma_1(P) &= -g^2 \mu^\epsilon
 \int_Q  D_0(Q+P)\, \Gamma_{45} \, G_0^{\rm T} (Q)\,  \Gamma_{45}\,,
 \text{ and } \\
 \label{rg8}
  \Pi_1(P) &=  g^2 \mu^\epsilon
 \int_Q  \Tr  \left [ \Gamma_{45}\, G_0(Q ) \,\Gamma_{45}  \,   G_0^{\rm T} (P-Q ) \right ] ,
\end{align}
respectively,
as can be seen in Fig.~\ref{self-en-fig}. Throughout the paper, we have denoted the integrals by:
\begin{align}
 \label{rg9} \int_Q =\int_{q_0} \int_{\vec{q}}  \, , \quad 
  \int_{q_0}  = \int  \frac{\mbox{d}q_0}{2\pi} \,,
 \quad \int_{\vec{q}} = \int  \frac{\mbox{d}^dq}{(2\pi)^d}\,.
\end{align}

Using Eq.~(\ref{rg15}), we get:
%%%%%%%%%%%%%%%%%%%%%%%%%%%%%%%%%%%%%%%%%%%%%%%%%%
\begin{align}
 \label{rg16} \Sigma_1(P) 
 &=-  g^2 \mu^\epsilon
 \int_Q  D_0(Q+P)\, G_0  (Q)
\nn  &=  g^2 \mu^\epsilon
 \int_Q \frac{ \left (   i\, q_0 + \xi\, q^2\right )\mathbb{1}_N 
-  \vec d (\vec{q})\cdot \vec \Gamma } 
 { \left [ \,  q_0^2 +\left (1- \xi^2\right )q^4 - 2\,i\, q_0\, \xi\,q^2 \right ]
  \left [ i\,  y \,\left( q_0 + p_0\right ) +  \left ( \mathbf q + \mathbf p \right )^2 \right ]} \,,\\
%%%%%%%%%%%%%%%%%%%%%%%%%%%%%%%%%%%%%%%%%%%%%  
 \label{rg17}\Pi_1(P) &=   g^2 \mu^\epsilon
\int_Q  \Tr    \left [G_0(Q)\, G_0(P-Q)\right ]  .
\end{align}
%%%%%%%%%%%%%%%%%%%%%%%%%%%%%%%%%

Performing the trace in the numerator of $\Pi_1$ yields
\begin{align}
 \nonumber   &\Tr    \left [ 
 \left \lbrace    i\, q_0 +\xi\, q^2-\vec d (\vec{q})\cdot \vec \Gamma \right\rbrace
  \left \lbrace    i\, \left(p_0-q_0\right ) +\xi\, \left (\vec p-\vec q\right )^2 
  -\vec d  \left (\vec p-\vec q\right ) \cdot \vec \Gamma \right\rbrace
 \right ]
\nn & = N \Bigl[ (-i\, q_0- \xi\, q^2)\Big \lbrace  i\left (q_0-p_0\right )
-\xi \left (\vec{q}-\vec{p}\right )^2\Big \rbrace+d_a(\vec{q}) \,d_a(\vec{q}-\vec{p})\Bigr]  \nn
%%%
&= N \Bigl[  (-i\, q_0- \xi\, q^2)\Big \lbrace  i\left (q_0-p_0\right )
-\xi \left (\vec{q}-\vec{p}\right )^2\Big \rbrace+
 \frac{4}{3} \big \lbrace \vec{q}\cdot(\vec{q}-\vec{p})\big \rbrace ^2-\frac{q^2}{3}(\vec{q}-\vec{p})^2 \Bigr]  \,,
\end{align}
where we have used Eq.~(\ref{d4}).
%%%%%%%%%%%%%%%%%%%%%%%%
Therefore,
\begin{align}
&\Pi_1(P) = 
\int_Q \frac{   g^2 \mu^\epsilon\, N \Bigl[  (-i\, q_0- \xi\, q^2)\Big \lbrace  i\left (q_0-p_0\right )
-\xi \left (\vec{q}-\vec{p}\right )^2\Big \rbrace+
 \frac{4}{3} \big \lbrace \vec{q}\cdot(\vec{q}-\vec{p})\big \rbrace ^2-\frac{q^2}{3}(\vec{q}-\vec{p})^2\Bigr] }
{  \left [ q_0^2 +\left (1- \xi^2\right )q^4 -2\,i\, q_0\, \xi\,q^2  \right ]\,
  \left [ \left (p_0-q_0\right )^2 +\left (1- \xi^2\right ) \left (\vec p-\vec q\right )^4 -2\,i \left (p_0-q_0\right ) \xi \left (\vec p-\vec q\right )^2  \right ] } \,.
%%%%
\end{align}

For the fermion self-energy, we first integrate over $q_0$, and then complete the 
squares in the denominator (using Feynman parametrization, if necessary) to make it easy to perform the $\mathbf q$-integral:
%%%%%%%%%%%%%%%%%%%%%%%%%%%%%
\begin{align}
\Sigma_1 (P) & =   \frac{ g^2 \mu^\epsilon} {2} 
 \int_{\vec q} 
 \frac{ 1  -  \frac{\mathbf d(\mathbf q) \cdot \mathbf \Gamma} {q^2}
 }
 {  i \,p_0\, y + q^2 \left ( 1 +y -\xi \, y \right )
+ p^2+2 \,\mathbf p \cdot \mathbf q }\nn
%%%%%%%%%%%%%%%%%%
& = -\frac{p^2 \,g^2 \,y  \left (1-\xi \right )
\left(\frac{\mu } {p}\right)^\epsilon }
{ 16 \,\pi ^2  \left (y+1 -\xi \, y \right )^3   \epsilon}
-
\frac{ i\, p_0\,
g^2 \,y\,   \left(\frac{\mu }{\sqrt{\left|p_0\right|}}\right)^\epsilon
}
{ 16\, \pi ^2 \left (y+1 -\xi \, y \right )^2   \epsilon}
+
\frac{ \left [  \mathbf d(\mathbf p) \cdot \mathbf \Gamma \right ]
g^2 \left(\frac{\mu }{p}\right)^\epsilon
}
{ 48 \,\pi ^2  \left (y+1 -\xi \, y \right )^3   \epsilon}
+ \mathcal{O} \left(  \epsilon^0\right),
\end{align}
where, in the second line, we have used the relation in Eqs.~(\ref{d4})~and~(\ref{ddint}).

For the boson self-energy, we integrate over $q_0$, employ 
partial fraction decomposition, complete the 
squares in the denominator (using Feynman parametrization, if necessary), before finally
integrating over $\mathbf q$:
\begin{align}
\Pi_1(P)& =  
\frac{
13\,i\,p_0 \, g^2\, N\,\xi 
\left(\frac{\mu } {\sqrt{\left|p_0\right|}}\right)^\epsilon}
{288 \,\pi ^2 \left(1-\xi ^2\right)^2 \epsilon }
-\frac{107\, p^2\,g^2 \,N  \left(\frac{\mu }{p}\right)^\epsilon}
{3456\, \pi ^2 \left(1-\xi ^2\right) \epsilon }
+ \mathcal{O} \left(  \epsilon^0\right).
\end{align}

%%%%%%%%%%%%%%%%
\subsection{RG equations}

The counterterm action is given by:
\begin{align}
 \label{rg1} 
 {\mathcal S}_{CT} = &  \int _P
\Big [  {\psi}^\dagger(P)
\left \lbrace  A_1\,i\,p_0 +A_2 \,d_a\left (\vec p \right )\,\Gamma_a  +  A_3\, {\xi} \, p^2  \right \rbrace  
{\psi}(P) 
  + {\phi}^*(P) \left (A_4 \, i\,{y}\,p_0+  A_5\, p^2 \right ) {\phi} (P)  \Big ]\nn
& 
+  {g}\,\mu^{\epsilon/2} \int_P \int_K
 A_6\left  [  {\phi} (P)\,{\psi}^\dagger (K) \,\Gamma_{45}\, {\psi}^*(P-K) 
+  {\phi}^* (P)\, {\psi}^{\rm T} (P-K)\,\Gamma_{45}\,  {\psi} (K)\right ]  , \nn
%%%%%%%%%%%%
A_{n} & \equiv \,Z_n-1 = 
\sum_{\lambda=1}^\infty \frac{Z_{ n,\lambda}}
{\epsilon^\lambda}  \text{ with }  n=1,2,3,4,5,6\,.
\end{align}
Adding the counterterms to the original $\mathcal S_0 $, and denoting the bare quantities by the index ``B'', we obtain the renormalized action as:
\begin{align}
 \label{ren-action} 
 {\mathcal S}_{ren}  & = \int _{P_B}
\Big [  {\psi}^\dagger(P_B)
\left \lbrace  i\,p_{0_B} + d_a\left (\vec p_B \right )\,\Gamma_a  +  {\xi}_B \, p_B^2  \right \rbrace  
{\psi}(P_B) 
  + {\phi}^*(P_B) \left ( i\,{y}_B\,p_{0_B}  +  p_B^2 \right ) {\phi} (P_B)  \Big ]\nn
& 
+ {g}_B  \int_{P_B} \int_{K_B}
 \left  [ {\phi} (P_B)\,{\psi}^\dagger (K_B) \,\Gamma_{45}\, {\psi}^*(P_B-K_B) 
+  {\phi}^* (P_B)\, {\psi}^{\rm T} (P_B-K_B)\,\Gamma_{45}\,  {\psi} (K_B)\right  ]  .
\end{align}
%%%%%%%%%%%%%%%%%%%%%%%

The bare and renormalized quantitites are related by the following convention:
%%%%%%%%%%%%%%%%%%%%%%
\begin{align}
\label{scale1}
&( p_0 )_B= Z_1 \, p_0\,,\quad 
\vec p_B = \sqrt{Z_2}\,\vec p \,, \quad  
\psi_B (P_B)= \sqrt {Z_\psi} \, \psi(P)\,,
\quad \phi_B ( P_B)= \sqrt {Z_\phi} \,  \phi( P)\,,\nn
& \xi_B= \frac{Z_3}{Z_2}\,\xi\,,\quad 
%%%%%%%%%%%%%5
y_B = \frac{Z_2\,Z_4} {Z_1\,Z_5}  \, y\,,\quad
%%%%%%%%%%%%
g_B= \frac{Z_6\, Z_2^{ 1 -\frac{d+2} {4}}}
{\sqrt{Z_1\,Z_5}}
\, g \,\mu^{\epsilon/2},
\quad Z_\psi= \frac{1} {Z_1\,Z_2^{\frac{d}{2}}} 
\,,\quad  Z_\phi=
\frac{Z_5 }
{Z_1\, Z_2^{ 1+\frac{d}{2}}} \,,
\end{align}
with the tree-level mass dimensions $[p_0]=2$, $[\vec p]=1$, $[\xi]=[y]=0\,.$

Let us define:
\begin{align}
z_\tau 
= 2  + \frac{\partial \ln {Z_1}} {\partial \ln \mu}  \,,\quad 
%%%%%%%%%%
z_r
= 1  + \frac{1} {2}\frac{\partial \ln {Z_1}} {\partial \ln \mu}  \,,\quad
%%%%%%%%%%%%%%%%%%%%%
\eta_\psi =\frac{1}{2} \frac{\partial \ln Z_\psi}{\partial \ln \mu}
  \,,\quad \eta_\phi =\frac{1}{2} \frac{\partial \ln Z_\phi}{\partial \ln \mu}\,,
\end{align}
where $z_\tau$ and $z_r$ are the quantum scaling dimensions for $k_0$ and $\mathbf k$, respectively.
Furthermore, $\eta_\psi$ and $\eta_\phi$ are the anomalous dimensions of the fermion and the boson, respectively, and can be expressed as:
\begin{align}
2\,\eta_\psi  &=
{2 + d - z_\tau - d \,z_r}
\,,
\quad 2\,\eta_\phi  =
4 + d - z_\tau - \left (2 + d \right ) z_r
+ \frac{1}{Z_5 } \frac{\partial  Z_5 }{\partial \ln \mu}\,.
\end{align}

Since the bare quantitites do not depend on $\mu$, their total derivative with respect to $\mu$ should vanish. Therefore, $\frac{d \ln g_B}{d \ln \mu}=0$ gives:
\begin{align}
-\frac{\partial g} {\partial \ln \mu} \equiv -\beta_g=
\left [ \frac{\epsilon}{2} 
+ \left(1-\frac{d}{2} \right) \left (z_r-1 \right )
-\frac{1}{2}   \frac{ \partial \ln Z_5}
{\partial \ln \mu }
+\frac{ \partial \ln Z_6} { \partial \ln \mu}
+ 1 - \frac{z_\tau} {2} 
\right ] \,g \,.
\end{align}

To one-loop order, we have $Z_n= 1+\frac{Z_{n,1}} {\epsilon}$, where
\begin{align}
 \label{rgmy} 
Z_{1,1} &= - \frac{g^2 \,y }
{ 16 \,\pi ^2 \left ( 1 +y -\xi\, y \right )^2} \,,\quad
%%%%%%%%%%
 Z_{2,1}  =  \frac{g^2}
 { 48 \,\pi ^2 \left ( 1 +y -\xi\, y \right )^3}  \,,\quad
%%%%%%%%%%%
Z_{3,1}  = - \frac{g^2 \,y  \left (1-\xi \right ) }
{  16 \,\pi ^2 \left ( 1 +y -\xi\, y \right )^3 \xi}  \,,\nn
 %%%%%%%%%
Z_{4,1}  &=   \frac{13 \,N\,g^2 \,\xi }
{288 \, \pi ^2 \left(1-\xi ^2\right)^2 y} \,,\quad
%%%%%%%%%%%%%%%%5
Z_{5,1} = -\frac{107 \,N\, g^2 }
{3456 \,\pi ^2 \left(1-\xi ^2\right)} \,,   \quad
 Z_{6,1} = 0 \,.
\end{align}

Using the expansions: 
\begin{align}
& z_\tau = z^{(0)}_\tau \,,\quad 
z_r = z^{(0)}_r \,,\quad
\eta_\psi= \eta_\psi^{(0)} + \epsilon\, \eta_\psi^{(1)}\,,\quad
\eta_\phi= \eta_\phi^{(0)} + \epsilon\, \eta_\phi^{(1)}\,,\nn
&
\beta_\xi=\beta_\xi^{(0)} +\epsilon \, \beta_\xi^{(1)}\,,
\quad \beta_y=\beta_y^{(0)} +\epsilon \, \beta_y^{(1)} \,,
\quad \beta_g=\beta_g^{(0)} +\epsilon \, \beta_g^{(1)} \,,
\label{expand0}
\end{align}
and comparing the powers of $\epsilon$ from the $\mu$-derivatives of the Eqs.~(\ref{scale1}), we get:
\begin{align}
&  z_\tau =2 
+ \frac{g^2\, y} {16 \,\pi ^2 \left (  1 +y -\xi\, y \right )^2}
\,,\quad
z_r = 1 -\frac{g^2} {96 \,\pi ^2 \left (  1 +y -\xi\, y \right )^3}
\,,\nn
%%%%%%%%
& \eta_\psi = - \frac{g^2\,\epsilon} {192 \, \pi ^2 \left (  1 +y -\xi\, y \right )^3}
+
\frac{g^2} {48\, \pi ^2 \left (  1 +y -\xi\, y \right )^3}
-\frac{g^2 \,y} {32 \,\pi ^2 \left (  1 +y -\xi\, y \right )^2}
\,,\nn
%%%%%%%%%%%%%
& \eta_\phi = -\frac{g^2\,\epsilon} {192 \, \pi ^2 \left (  1 +y -\xi\, y \right )^3}
+
\frac{g^2}{\pi ^2} \left[ 
\frac{107 \,N} {6912 \left(1-\xi ^2\right)}
-\frac{y} {32 \left (  1 +y -\xi\, y \right )^2}
+\frac{1} {32 \left (  1 +y -\xi\, y \right )^3}
\right ]\,,
\end{align}
and the beta-functions:
%%%%%%%%%%%%%%%%%%
\begin{align}
& \beta_\xi = 
-\frac{\xi +3 \left (1-\xi \right ) y} 
{48\, \pi ^2 \left ( 1 +y -\xi\, y \right )^3}\,g^2 \,, \nn &
%%%%%%%%%%%%%%%%%%%%%%%%
  \beta_y  =  \frac{g^2} {\pi ^2}
\left[
\frac{13\,N \,\xi }
{288 \left(1-\xi ^2\right)^2}
+\frac{107 \,N\, y}
{3456 \left(1-\xi ^2\right)}
+\frac{y^2} {16 \left ( 1 +y -\xi\, y \right )^2}
+\frac{y} {48 \left ( 1 +y -\xi\, y \right )^3}
\right ]  \,,\nn &
%%%%%%%%%%%%%%%%%%%%%%%%%%%%%%%%%
\beta_g =- \frac{\epsilon \,g }  {2} 
+ 
\frac{g^3} {\pi ^2}
 \left[ 
 \frac{107\, N } {6912 \left(1-\xi ^2\right)}
 +\frac{y} {32 \left ( 1 +y -\xi\, y \right )^2}
 -\frac{1} {96 \left ( 1 +y -\xi\, y \right )^3} \right ].
\end{align}

%%%%%%%%%%%%%%%%%%%%%%%%%%%%%%%%%%%%%%%%%%%%%%%%%%%%%%
\subsection{Fixed points and their stability}

The fixed points $\left ( \xi^*, y^*, {g}^* \right )$ are given by:
\begin{align}
\left (0, \,0, \,0 \right ) \text{ and }
\left (0,\, 0, \,
\pm \frac{24 \,\sqrt{6}\, \pi \, \sqrt{\epsilon} }
{\sqrt{ 107 \,N -72 }}
 \right ) .
\end{align}

%%%%%%%%%%%%%%%%%%%%%%%%%%
\begin{figure}
\centering
\subfigure[]{\includegraphics[width = 0.4 \columnwidth]{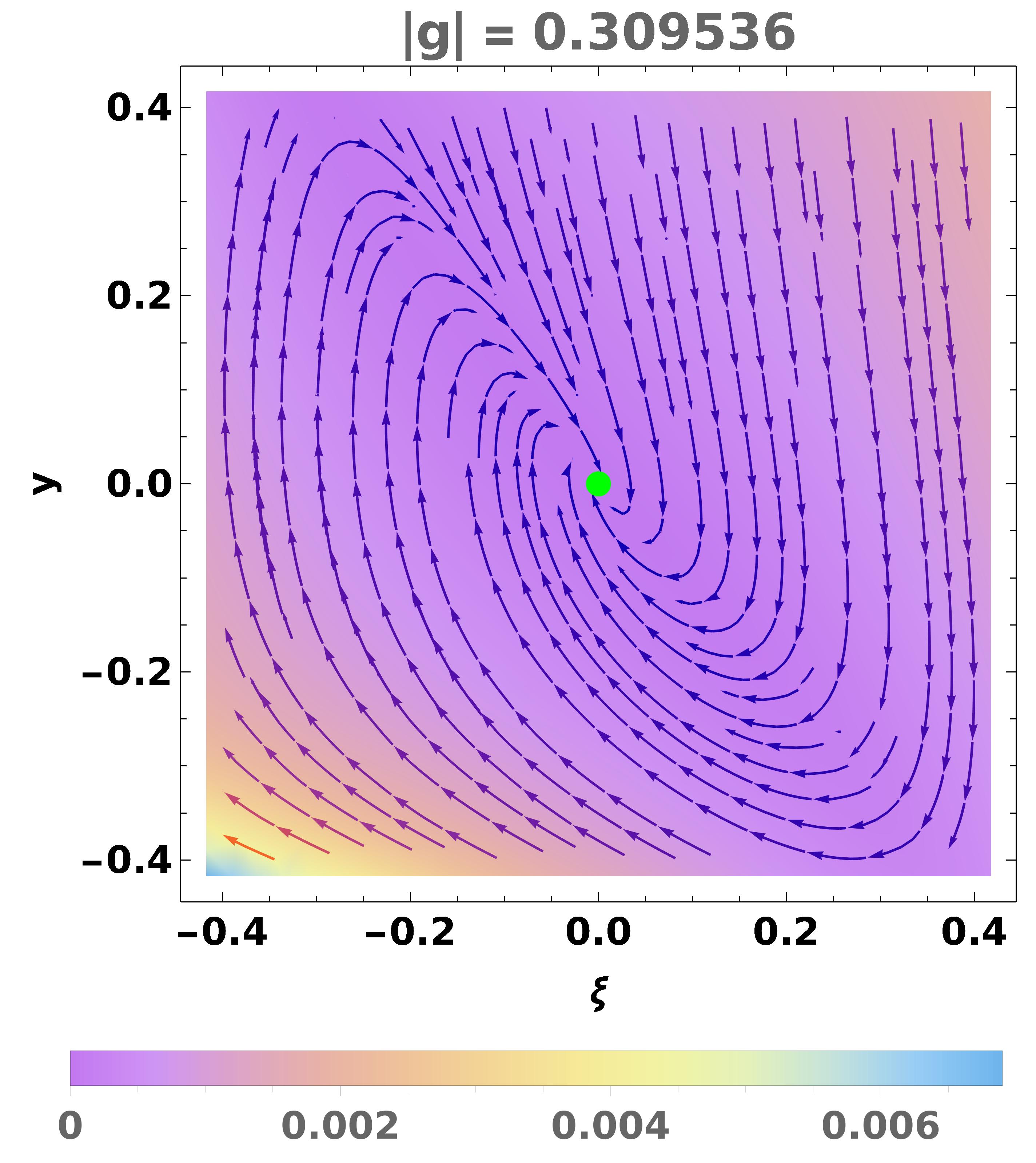}} \hspace{1 cm}
\subfigure[]{\includegraphics[width = 0.4 \columnwidth]{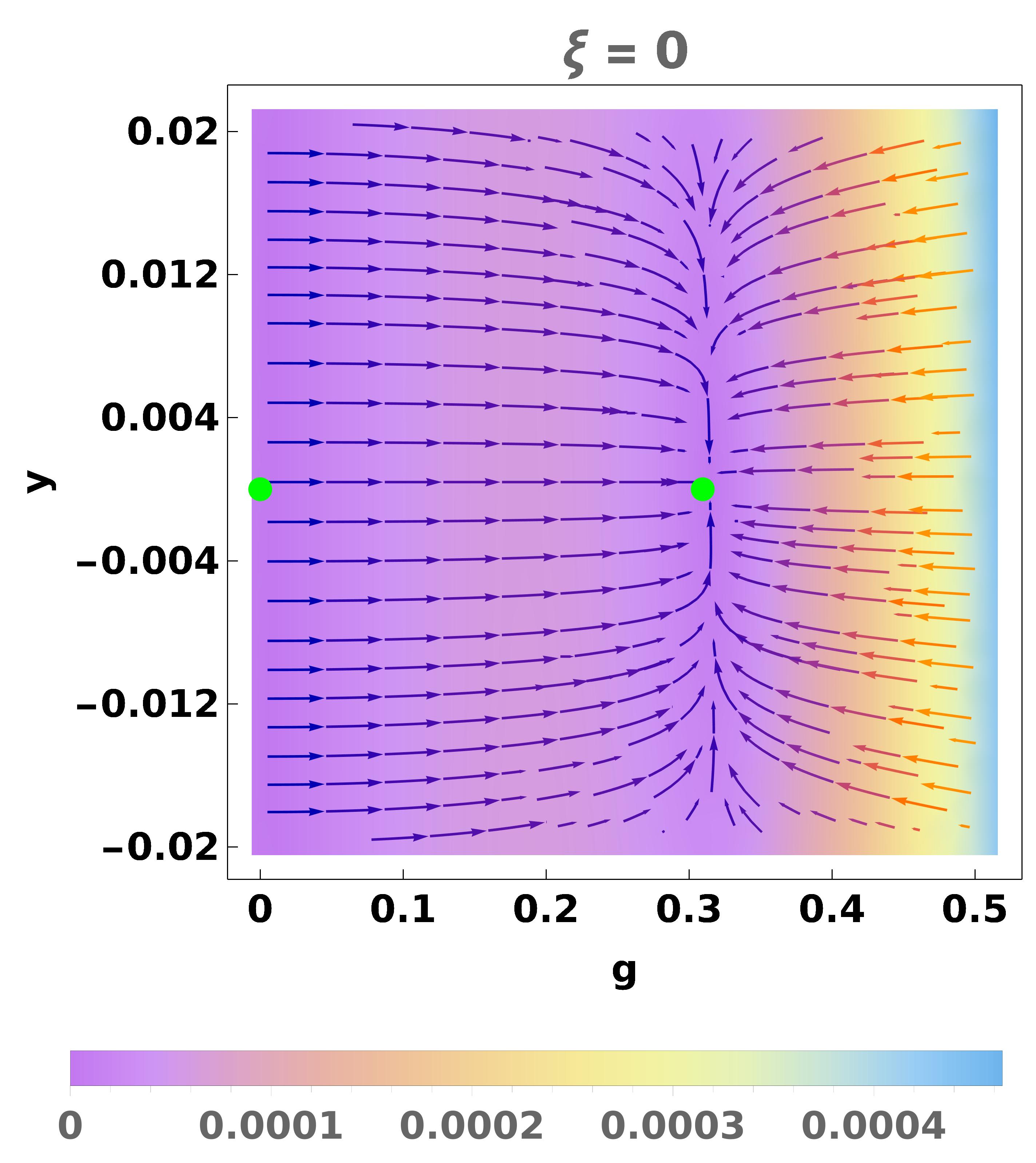}} 
%\subfigure[]{\includegraphics[width = 0.3 \columnwidth]{flow3}}
\caption{\label{clean-fig}
The RG flow diagrams for the clean system (no disorder) for $N=4$ and
$\epsilon= 0.001$.
The green dots show the positions of the RG fixed points.}
\end{figure}
%%%%%%%%%%%%%%%%%%%%%%%%

To analyze the stability of the non-Gaussian fixed point at
$ |g^*|= \frac{24 \,\sqrt{6}\, \pi \, \sqrt{\epsilon} }
{\sqrt{ 107 \,N -72}}$, we write down the linearized flow equations in its vicinity, which are:
\begin{align}
\frac{d}{d l}
\left(\begin{array}{c} 
\delta \xi\\ \delta y \\ \delta  g   
\end{array}\right) \Bigg\rvert _{(  \xi^*, \,   y^*,\, g^*)}
 \approx 
\mathcal M
\left(\begin{array}{c} 
\delta \xi\\ \delta y \\ \delta  g 
 \end{array}\right).
\end{align}
Here $l = - \ln \mu $ is the logarithmic length scale determining the RG flows towards the infrared, and
\begin{align}
\mathcal M =
%%%%%%%%%
\frac{\epsilon} {  107\,N -72 }
\left(
\begin{array}{ccc}
 72 & 216 & 0 \\
 -156 \,N & -107 \,N-72 & 0 \\
 0 & -\frac{5184 \,\sqrt{6}\, \pi \, \sqrt{\epsilon }}
 {\sqrt{107\,N-72}} & 72-107\,N \\
\end{array}
\right) \,.
%%%%%%%%%%%   
\end{align}
The eigenvalues of $\mathcal M$ are given by:
\begin{align}
\left( -   \epsilon ,\,
-\frac{\epsilon }{2} 
-\frac{36 \pm
\frac{\sqrt{11449 \,N^2-103968 \,N+20736} } {2}
}
{107 \,N-72}\, \epsilon
 \right).
\end{align}
%%%%%%%%%%%%%%%%%%%%%%%%%%%
For $N=4$ or $N=8$, the last two eigenvalues are complex conjugates of each other, and have negative real parts. The nonzero imaginary parts of the pair of complex eigenvalues (with negative real parts) indicate that the fixed point is a  stable (or attractive) limit cycle in the $y-\xi$-plane [cf. Fig.~\ref{clean-fig}(a)]. For $N >8 $, all the eigenvalues are real and negative.
These results show that this non-Gaussian fixed point is stable in the infrared. 
On the other hand, the Gaussian fixed point is unstable, as the eigenvalues of the stability matrix are $\left(\epsilon/2, \,0, \,0 \right)$.
Two representative RG flow diagrams are shown in Fig.~\ref{clean-fig}.

%%%%%%%%%%%%%%%%%%%%%%%%%%%%%%%%%%%%%%%%%%%%%%%%%%%%%%%%%%%%%%%%%%%%%%

\section{Effect of short-range-correlated disorder}
\label{disorder}

In this section, we consider the effect of adding disorder to the system of the form:
\begin{align}
H_{dis}= \int d^d x\,\Big [ V_0 (x)\, \left( \psi^{\dag} \, {\psi}\right )_{\tau}
+ V_1 (x)\, \sum \limits_{a}\left( \psi^{\dag} \,\Gamma_a\, {\psi}\right )_{\tau} 
+ + V_2 (x)\, \sum \limits_{a< b }\left( \psi^{\dag} \,\Gamma_{ab} \, {\psi}\right )_{\tau} \Big ]\,,
\label{H_dis}
\end{align}
where $V_\alpha(x)$'s are produced by impurities or defects with and without spinor structure.
We consider the case of short-range-correlated disorder such that
\begin{align}
\langle V_\alpha (x) \, V_{ \alpha'}   (x')\rangle _{avg}= W_{\alpha} \, \delta(x-x')\, \delta_{\alpha \alpha'}\,.
\label{disav}
\end{align}
The parameter $W_\alpha$ measures the strength of the disorder induced by the distribution of impurities. 
We introduce $n$ copies of the fields $\psi \to \psi_{i} $ with $ i \in [1,n ]$, and average over the disorder using Eq.~(\ref{disav}). 
This is the standard treatment of disorder in the replica formalism, where the number of replicas $n \rightarrow 0$ at the end of the computation.
The replica term in the action is then given by:
\begin{align}
S_{\text{dis}} =
& - W_0\,\mu^\epsilon   \sum \limits_{i,j=1 } ^n  \int d\tau\, d\tau' \,d^d x \,( \psi_i^{\dag} \, {\psi_i})_{\tau} \,( \psi_j^{\dag} \,  \psi_j  )_{\tau'} 
- W_1 \,\mu^\epsilon \sum  \limits_{i,j=1}^n 
\sum \limits_{a}  \int d\tau \,d\tau' \,d^d x\, ({ \psi_i}^{\dag}\, \Gamma_a^i \, {\psi_i})_{\tau} \,( \psi_j^{\dag} \,\Gamma_a^j \,  \psi_j )_{\tau'}\nn
 &- W_2 \,\mu^\epsilon  \sum  \limits_{i,j=1 }^n  
 \sum \limits_{a<b}\int d\tau \,d\tau' \,d^d x\, ({ \psi_i}^{\dag} \, 
 \Gamma_{ab}^i \, {\psi_i})_{\tau} \,( \psi_j^{\dag}\, \Gamma_{ab}^j\,  \psi_j)_{\tau'}\,, \label{disorderaction}
\end{align}
where $i$ and $j$ denote the replica indices.
The tree-level mass dimension of $W_{\alpha}$ is
$ [W_{\alpha}] = 4 - d   $.
Let us denote the corresponding matrices by $M_\alpha$, such that
\begin{align}
\left ( M_0,M_1, M_2 \right) = (\mathbb{1}_{N_c}, \Gamma_ a,  \Gamma_ {ab})\,,
\end{align} 
where $N_c$ is the dimensionality of the gamma matrices. For the current problem, $N_c=5$.

%%%%%%%%%%%%%%%%%%%%
\subsection{Fermion self-energy correction from disorder}
\label{disorderself}

\begin{figure}
\includegraphics[width = 0.27 \columnwidth]{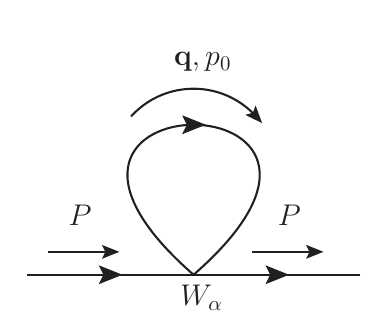}
\caption{\label{self-dis-fig} One-loop correction to fermion self-energy from a disorder vertex.}
\end{figure}

A sample contraction of terms leading to the one-loop correction to fermion self-energy from a disorder vertex looks like:
\begin{align}
   \left( \psi^\dagger \, M \,\psi \right )   \left( \psi^\dagger \, M \,\psi \right )
& = \left( \psi^*_\gamma \, M^{\gamma,\delta} \psi_\delta \right )   
\left( \psi^*_\lambda \, M^{\lambda, \sigma } \psi_\sigma \right ) 
%%%%%%%%%%%%%%
= \psi^*_\gamma \left[ M^{\gamma,\delta} \, \left (G_0\right )_{\delta, \lambda } \, M^{\lambda,\sigma} \right ] \psi_\sigma\,.\nonumber
\end{align}
%%%%%%%%%%%%%%%%%%%%%%%%%%%%%%%%%%%%%%%
Therefore, the one-loop fermion self-energy correction from disorder, shown in Fig.~\ref{self-dis-fig}, is given by the term:
\begin{align}
\Sigma_{1d} (P)
&= - 2\,\mu^\epsilon
\left [ W_0 \int_{\vec q} G_0(\vec q , p_0 )
+ W_1 \sum_a \int_{\vec q} \Gamma_a\, G_0(\vec q, p_0) \,\Gamma_a
+ W_2 \sum_{a<b} \int_{\vec q} \Gamma_{ab}\, G_0( \vec q, p_0) \,\Gamma_{ab}
\right ]
\nn & = - 2 \,\mu^\epsilon  \left [ W_0  + N_c\, W_1 + \frac{N_c \left(N_c-1\right)}{2}\right ]
\int_{\vec q}\,  \frac{-i\, q_0 -\xi\,q^2  }{p_0^2  +  \left(1-\xi^ 2 \right) q^4 -2\, i\, p_0 \,\xi \,q^2  }
\nn & = 
\frac{\left( 1+\xi^2 \right) i\,p_0 \left [ W_0  + N_c\, W_1 + \frac{N_c \left(N_c-1\right)}{2}\right ]}
{4\,\pi^2  \left(1-\xi^2 \right)^2 \epsilon}  \left( \frac{\mu}{ \sqrt{|p_0|} }\right)^\epsilon 
+ \mathcal{O} \left(  \epsilon^0\right) ,
\end{align}
using Eqs.~(\ref{rel0}) and (\ref{rel1}).

%%%%%%%%%%%%%%%%%%%%
\subsection{Fermion-boson vertex correction from disorder}
\label{disordervert}

\begin{figure}
\centering
\subfigure[]{\includegraphics[width = 0.37 \columnwidth]{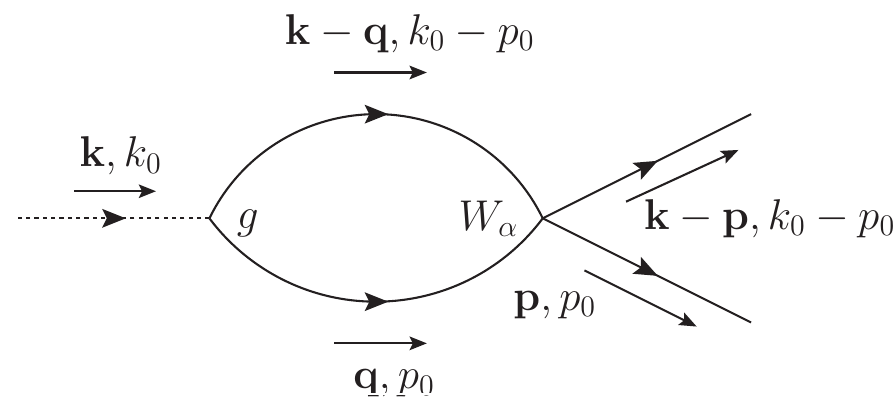}} \hspace{2 cm}
\subfigure[]{\includegraphics[width = 0.37\columnwidth]{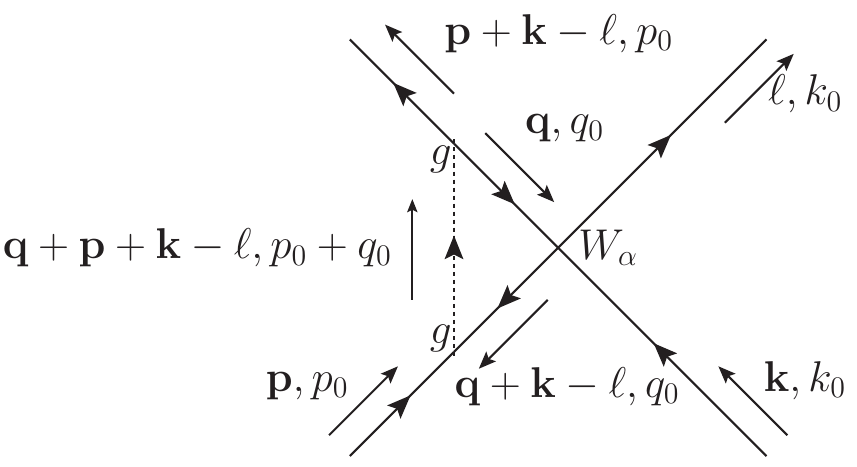}}
\caption{\label{disorderloops} One-loop correction to (a) fermion-boson vertex from disorder, and (b) disorder from fermion-boson vertex.}
\end{figure}

Let us compute the one-loop fermion-boson vertex correction coming from the disorder vertices.
A sample contraction of terms leading to this looks like:
\begin{align}
\phi^* \left( \psi^T \, \Gamma_{45} \,\psi  \right )  \left( \psi^\dagger \, M \,\psi \right )   \left( \psi^\dagger \, M \,\psi \right )
& = \phi^* \left( \psi_\alpha \left ( \Gamma_{45}\right )^{\alpha, \beta} \, \psi_\beta \right )  \left( \psi^*_\gamma \, M^{\gamma,\delta} \psi_\delta \right )   
\left( \psi^*_\lambda \, M^{\lambda, \sigma } \psi_\sigma \right ) 
%%%%%%%%%%%%%%
\nn & = \phi^* \, \psi_\delta \left[   \left ( G_0\right)_{\alpha, \gamma} \, M^{\gamma,\delta}\left ( \Gamma_{45}\right )^{\alpha, \beta} \,  \left ( G_0\right)_{\beta,\lambda} \, M^{\lambda, \sigma }
\right]  \psi_\sigma
%%%%%%%%%%%%%%
\nn & = \phi^* \, \psi_\delta \left[   \left( M^T\right)^{\delta,\gamma} \left(   G_0 ^T\right)_{\gamma,\alpha}
\left ( \Gamma_{45}\right )^{\alpha, \beta} \,  \left ( G_0\right)_{\beta,\lambda} \, M^{\lambda, \sigma }
\right]  \psi_\sigma\,.\nonumber
%%%%%%%%%%%%%%%%%%%%%%%
\end{align}
or
%%%%%%
\begin{align}
\phi \left( \psi^\dagger \, \Gamma_{45} \,\psi^*  \right )  \left( \psi^\dagger \, M \,\psi \right )   \left( \psi^\dagger \, M \,\psi \right )
& = \phi \left( {\psi^{\dagger}}_\alpha \left ( \Gamma_{45}\right )^{\alpha, \beta} \, 
\psi^*_\beta \right )  \left( \psi^*_\gamma \, M^{\gamma,\delta} \psi_\delta \right )   \left( \psi^*_\lambda \, M^{\lambda, \sigma } \psi_\sigma \right ) 
%%%%%%%%%%%%%%
\nn & = \phi \, \psi^*_\gamma \left[   M^{\gamma,\delta}   \left ( G_0\right)_{\delta,\alpha } \left ( \Gamma_{45}\right )^{\alpha, \beta} \,  \left ( G_0\right)_{\sigma,\beta } \, M^{\lambda, \sigma }
\right]  \psi^*_\lambda
%%%%%%%%%%%%%%
\nn & = \phi \,  \psi^*_\gamma \left[   M^{\gamma,\delta}  
\left(   G_0 \right)_{\delta,\alpha }
\left ( \Gamma_{45}\right )^{\alpha, \beta} \,  \left ( G_0^T \right)_{\beta,\sigma } \, \left( M^T\right)^{ \sigma,\lambda }
\right]  \psi^*_\lambda \,.\nonumber
%%%%%%%%%%%%%%%%%%%%%%%
\end{align}
Therefore, we need to evaluate loop integrals of the form:
\begin{align}
I(\alpha) &= -2\, g\, W_\alpha \, \mu^{\frac{3\,\epsilon}{2}} 
\int_{\vec q} M_\alpha\, G_0(\vec k - \vec q, k_0-p_0)  \, \Gamma_{45} \, G_0^T(\vec q, p_0)\,M_\alpha^T \nn
&= -2\, g\, W_\alpha \, \mu^{\frac{3\,\epsilon}{2}} 
\int_{\vec q} M_\alpha\, G_0(\vec k - \vec q, k_0-p_0)   \, G_0(\vec q, p_0)  \, \Gamma_{45} \,M_\alpha^T \nn
%%%%%%%%%%%%%%%%
&= -2\, g\, W_\alpha \, \mu^{\frac{3\,\epsilon}{2}}  \int_{\vec q} \frac{  M_\alpha
\left[ i\left ( p_0-k_0\right ) -\xi\, \left( \vec q-\vec k\right)^2 + \mathbf d\left( \vec q-\vec k \right) \cdot \mathbf  \Gamma \right ]\,
\left[- i \,p_0-\xi\,   q^2 + \mathbf d  \left ( \mathbf q  \right ) \cdot \mathbf  \Gamma \right ]  \Gamma_{45} \, M^T_\alpha
}
{ \left [ \left (p_0-k_0\right ) ^2 +\left(1-\xi^2\right ) \left ( \mathbf q -\mathbf k\right ) ^4 +2\, i \left (p_0-k_0\right ) \,\xi\,\left ( \mathbf q -\mathbf k\right )^2 \right ] 
\left [  \, p_0^2 +\left(1-\xi^2\right ) q^4 -2\, i \,p_0 \,\xi\,q^2 \, \right ] 
}\,,
\end{align}
corresponding to the one-loop vertex correction figure shown in Fig.~\ref{disorderloops}(a).
We set $\vec k = k_0 =0 $ without any loss of generality, as this will still allow us to extract the divergent part.

Using Eq.~(\ref{rg15}), for $\alpha=(0,1) $, we get:
\begin{align}
\label{i1}
I(\alpha) 
& =- 2\, g\, W_\alpha \, \mu^{\frac{3\,\epsilon}{2}}  \int_{\vec q} \frac{  M_\alpha
\left[ i \, p_0 -\xi\, q^2 + \mathbf d\left( \vec q-\mathbf k \right) \cdot \mathbf  \Gamma \right ]\,
\left[- i \,p_0-\xi\,   q^2 + \mathbf d  \left ( \mathbf q  \right ) \cdot \mathbf  \Gamma \right ]  M_\alpha
}
{ \left [  p_0^2 +\left(1-\xi^2\right ) q ^4 + 2\, i \,p_0 \,\xi\,\left ( \mathbf q -\mathbf k\right )^2 \right ] 
\left [  \, p_0^2 +\left(1-\xi^2\right ) q^4 -2\, i \,p_0 \,\xi\,q^2 \, \right ] 
}  \, \Gamma_{45}\,.
\end{align}
Using Eq.~(\ref{rg151}), we get:
\begin{align}
\label{i2}
I(2) &= 2\, g\, W_2\, \mu^{\frac{3\,\epsilon}{2}}  \int_{\vec q} \frac{  M_2
\left[ i \, p_0 -\xi\, \left( \vec q-\vec k\right)^2 + \mathbf d\left( \vec q-\vec k \right) \cdot \mathbf  \Gamma \right ]\,
\left[- i \,p_0-\xi\,  q^2 + \mathbf d  \left ( \mathbf q  \right ) \cdot \mathbf  \Gamma \right ] M_2
}
{ \left [  p_0^2 +\left(1-\xi^2\right ) \left ( \mathbf q -\mathbf k\right ) ^4 +2\, i \,p_0 \,\xi\,\left ( \mathbf q -\mathbf k\right )^2 \right ] 
\left [  \, p_0^2 +\left(1-\xi^2\right ) q^4 -2\, i \,p_0 \,\xi\,q^2 \, \right ] 
}  \, \Gamma_{45}\,.
\end{align}

Let us now compute the integral
\begin{align}
J &\equiv  \int_{\vec q} \frac{  
\left[ i\, p_0  -\xi\, q^2 + \mathbf d\left( \vec q  \right) \cdot \mathbf  \Gamma \right ]\,
\left[- i \,p_0-\xi\,   q^2 + \mathbf d  \left ( \mathbf q  \right ) \cdot \mathbf  \Gamma \right ]  
}
{ \left [ p_0 ^2 +\left(1-\xi^2\right ) q ^4 +2\, i \,p_0 \,\xi\,q^2 \right ] 
\left [  \, p_0^2 +\left(1-\xi^2\right ) q^4 -2\, i \,p_0 \,\xi\,q^2 \, \right ] 
}  \,.
\end{align}
Dropping the terms which do not contribute to vertex correction, we are left with:
\begin{align}
J ' &\equiv  \int_{\vec q} \frac{  
-\left( i\,p_0 +\xi\, q^2 \right )
\left(i \,p_0 - \xi\,  q^2  \right ) 
+    \vec d ^ 2(\mathbf q)
}
{ \left [  \,p_0^2 +\left(1-\xi^2\right ) q^4\right ]^2  + 4 \,p_0^2 \,\xi^2\,q^4  } 
=  \frac{ \left( 1+\xi^2 \right) |p_0|^ {  -\frac{ \epsilon}{2} }}
{8\,\pi^2 \left( 1-\xi^2 \right) \epsilon}
+ \mathcal{O} \left(  \epsilon^0\right).
\end{align}

Hence, using Eqs.~(\ref{rel0}) and (\ref{rel1}) in Eqs.~(\ref{i1}) and (\ref{i2}), the divergent part of the fermion-boson vertex correction from disorder is given by:
\begin{align}
\Gamma_V= -\frac{ g\, \mu^{\frac{ \epsilon}{2}}\left( 1+\xi^2 \right)  
\left [  W_0 +  N_c \, W_1  -\frac{N_c \left( N_c-1 \right) W_2 }{2} \right ]}
{ 4\,\pi^2 \left( 1-\xi^2 \right) \epsilon}  \left( \frac{\mu}{ \sqrt{|p_0|} }\right)^\epsilon\,.
\end{align}

%%%%%%%%%%%%%%%%%%%%%%%%%%
\subsection{Corrections to disorder vertices from fermion-boson vertex}
\label{verttodisorder}

Fig.~\ref{disorderloops}(b) shows how the fermion-boson vertex can generate a one-loop correction for each disorder vertex.
A sample contraction of terms leading to this looks like:
\begin{align}
&\phi^* \left( \psi^T \, \Gamma_{45} \,\psi  \right ) \phi\left( \psi^\dagger \, \Gamma_{45} \,\psi ^* \right ) 
 \left( \psi^\dagger \, M \,\psi \right )   \left( \psi^\dagger \, M \,\psi \right )
= \phi\, \phi^* \left[ \psi_\alpha \left ( \Gamma_{45}\right )^{\alpha, \beta} \, \psi_\beta \right ] 
\left[ \psi^*_\gamma \left ( \Gamma_{45}\right )^{\gamma, \delta} \, \psi^*_\delta \right ]
 \left( \psi^*_\lambda \, 
M^{\lambda,\sigma} \psi_\sigma \right )   
 \left( \psi^\dagger \, M \,\psi \right )
%%%%%%%%%%%%%%
\nn & = 
D_0 \left[ \psi_\alpha \left ( \Gamma_{45}\right )^{\alpha, \beta} \, \psi_\beta \right ] 
 \left( \psi^*_\lambda \, 
M^{\lambda,\sigma} \psi_\sigma \right )  
\left[ \psi^*_\gamma \left ( \Gamma_{45}\right )^{\gamma, \delta} \, \psi^*_\delta \right ] 
 \left( \psi^\dagger \, M \,\psi \right )
%%%%%%%%%%%%%%%%%%%%%%%%%%% 
\nn & = 
D_0\left[  \psi_\alpha \left ( \Gamma_{45}\right )^{\alpha, \beta} \, 
\left( G_0 \right )_{\beta ,\lambda}
M^{\lambda,\sigma} \left( G_0 \right )_{\sigma,\gamma} 
\left ( \Gamma_{45}\right )^{\gamma, \delta} \, \psi^*_\delta \right]
 \left( \psi^\dagger \, M \,\psi \right )
%%%%%%%%%%%%%%%%%%%%%%%%%%%%%%%%%%%%%
\nn & = - \left[ \psi^*_\delta \left \lbrace D_0\, 
\left (\Gamma_{45} \right) ^{\delta, \gamma} \left(   G_0 ^T\right)_{\gamma,\sigma}
\left( M^T\right)^{\sigma, \lambda} \left(   G_0 ^T\right)_{\lambda, \alpha}
\left ( \Gamma_{45}\right )^{\alpha, \beta} 
\right \rbrace  \psi_\beta  \right]
\left( \psi^\dagger \, M \,\psi \right )  .\nonumber
\end{align}
There are eight such terms.

%Another sample contraction of terms leading to this looks like:
%\begin{align}
%&\phi^* \left( \psi^T \, \Gamma_{45} \,\psi  \right ) \phi\left( \psi^\dagger \, \Gamma_{45} \,\psi ^* \right ) 
% \left( \psi^\dagger \, M \,\psi \right )   \left( \psi^\dagger \, M \,\psi \right )
%= \phi\, \phi^* \left[ \psi_\alpha \left ( \Gamma_{45}\right )^{\alpha, \beta} \, \psi_\beta \right ] 
%\left( \psi^*_\lambda \, 
%M^{\lambda,\sigma} \psi_\sigma \right )
%%%%%%%%%%%%%%%
% \left( \psi^*_\rho \, M^{\rho,\nu} \,\psi_\nu \right )
%\left[ \psi^*_\gamma \left ( \Gamma_{45}\right )^{\gamma, \delta} \, \psi^*_\delta \right ]   
%%%%%%%%%%%%%%
%\nn & = D_0 \left[ \psi_\alpha \left ( \Gamma_{45}\right )^{\alpha, \beta} \, 
%(G_0)_{\beta,\lambda} \, 
%M^{\lambda,\sigma} \psi_\sigma \right ]
%%%%%%%%%%%%%%%
% \left[ \psi^*_\rho \, M^{\rho,\nu} \,
% (G_0)_{\nu ,\gamma} \left ( \Gamma_{45}\right )^{\gamma, \delta} \, \psi^*_\delta \right ]   
%.\nonumber
%\end{align}
%This BCS or ZS' diag does not correct disorder vertex.

Using Eqs.~(\ref{rg15}) and (\ref{rg151}), The contributions will be given by integrals of the form:
\begin{align}
&  4 \,(-1)^2\, W_\alpha\,g^2 \,\mu^{2\,\epsilon} \int_Q \Gamma_{45}\, G_0^T(Q) \, M_\alpha^T \,
G_0^T (\vec q+\vec k -\boldsymbol \ell, q_0)\,\Gamma_{45}\, 
D_0 \left( \vec q +\vec p +\vec k-\boldsymbol \ell, p_0 +q_0 \right) \nn
& = 4\,\Omega_\alpha\, W_\alpha\,g^2 \,\mu^{2\,\epsilon} \int_Q   G_0 (Q) \, M_\alpha \,
G_0 (\vec q+\vec k -\boldsymbol \ell, q_0) \,
 D_0 \left( \vec q +\vec p +\vec k- \boldsymbol \ell, p_0 +q_0 \right) \,,
\end{align}
where $\Omega_\alpha =1$ for $\alpha=0$ and $\alpha=1$, while $\Omega_\alpha =-1$ for $\alpha=2$.
For extracting the divergent parts, we can set $ \vec p =\vec k =\boldsymbol \ell =0$.

Using Eqs.~(\ref{rel0}), (\ref{rel1}), (\ref{rel2}) and (\ref{ddint}), the contributions for scalar, vector, and tensor disorders reduce to: 
\begin{align}
& J_s =  4\, W_0 \,g^2 \,\mu^{2\,\epsilon}  \int_Q
\frac{ \left ( i\, q_0 +\xi \, q^2 \right )^2  +q^4}
{ \left [ q_0^2 +     \left( 1-\xi^2 \right) q^4-2\, i\, q_0\,\xi \,q^2 
\right ] ^2  
\left[  i\, y \left( q_0 + p_0 \right )+q^2 \right ]}  \,,\\
%%%%%%%%%%%%%%%%%%%%%%
& J_v =  4\, W_1 \,g^2 \,\mu^{2\,\epsilon}  \int_Q
\frac{ \left ( i\, q_0 +\xi \, q^2 \right )^2  -  \frac{\left( N_c-2 \right) q^4}{N_c } }
{ \left [ q_0^2 +     \left( 1-\xi^2 \right) q^4-2\, i\, q_0\,\xi \,q^2 
\right ] ^2  \left[  i\, y \left( q_0 + p_0 \right )+q^2 \right ]}  \,, \\
%%%%%%%%%%%%%%%%%%%%%
%%%%%%%%%%%%%%%%%%%%%%
& J_t = - 4\, W_2 \,g^2 \,\mu^{2\,\epsilon}  \int_Q
\frac{ \left ( i\, q_0 +\xi \, q^2 \right )^2  +  \frac{\left( N_c-4 \right) q^4}{N_c } }
{ \left [ q_0^2 +     \left( 1-\xi^2 \right) q^4-2\, i\, q_0\,\xi \,q^2 
\right ] ^2  \left[  i\, y \left( q_0 + p_0 \right )+q^2 \right ]}  \,.
\end{align}

On performing the integrals, we get:
\begin{align}
& J_s = \frac{g^2 \, W_0\,\mu^{\epsilon} \, y}{4\, \pi^2 \left( 1+ y -\xi\, y\right)^2\, \epsilon }
 \left( \frac{\mu}{ \sqrt{|p_0|} }\right)^\epsilon + \mathcal{O} \left(  \epsilon^0\right),
\\ & J_v = \frac{g^2 \, W_1\,\mu^{\epsilon} \, \left[ 1+ 2 \, y-N_c  \,(1+y) -(N_c-1) \, \xi\, y   \right ]  }
{4\, \pi^2 \, N_c\left( 1+ y -\xi\, y\right)^2\, \epsilon }
 \left( \frac{\mu}{ \sqrt{|p_0|} }\right)^\epsilon + \mathcal{O} \left(  \epsilon^0\right),
 \\ & J_t = \frac{g^2 \, W_2\,\mu^{\epsilon} \, \left[ 2- \, y\,(N_c -4 + 2\, \xi )  \,\right ]}
{4\, \pi^2 \, N_c\left( 1+ y -\xi\, y\right)^2\, \epsilon }
 \left( \frac{\mu}{ \sqrt{|p_0|} }\right)^\epsilon + \mathcal{O} \left(  \epsilon^0\right),
\end{align}
for the scalar, vector, and tensor disorder vertices, respectively.

%%%%%%%%%%%%%%%%%%%%%%%%%%%%%%%
\subsection{Corrections to disorder vertices from themselves}
\label{distodisorder}

The loop corrections to the disorder lines themselves come from the fully connected contractions of
\begin{align}
\label{disordercontraction}
\delta S_{\text{dis}} & = \frac{1}{2} \int  d\tau\, d\tau'\,  d\tau''\, d\tau'''\,
d^d \mathbf x\, d^d \mathbf x' 
   \sum \limits_{i,j,k,l =1}^n \mathcal{C}_{ijkl}\,, \text{ where}\nn
\mathcal{C}_{ ijkl } & =
 \sum \limits_{\alpha,\alpha' = 0}^2 
  W_{\alpha} \,W_{\alpha'}
\left  ( \psi_i^{\dag}\,M^i_{\alpha}\,  {\psi_i } \right)_{\mathbf x,\tau}
\left  ( \psi_j^{\dag} \,M^j_{\alpha}\, \psi_j \right)_{\mathbf x,\tau'} 
\left  ( \psi_k ^{\dag} \,M^k_{\alpha'}\, \psi_k  \right )_{\mathbf x',\tau''} 
\left( {\psi _l}^{\dag}\,M^l_{\alpha'}\, \psi_l \right )_{\mathbf x',\tau'''} \,,
\end{align}
where we have kept track of the replica indices on the $\Gamma$-matrices.
We need to consider three types of diagrams, as enumerated below:
%%%%%%%%%%%%%%%%%
\begin{enumerate}

\item

A sample contraction of terms leading to a ZS diagram from disorder vertices takes the form:
\begin{align}
& W_{{\alpha} }\, W_{{\alpha'}}\sum \limits_{i,j,k,l} 
 \left( \psi^\dagger_i \, M^i_\alpha \,\psi_i \right )_{\mathbf x,\tau} 
  \left( \psi^\dagger_j \, M^j_\alpha \,\psi_j \right )_{\mathbf x,\tau'} 
\left( \psi^\dagger_k \, M^k_{\alpha'} \,\psi_k \right )_{\mathbf x',\tau''}    
  \left( \psi^\dagger_l \, M^l_{\alpha'} \,\psi_l \right )_{\mathbf x',\tau'''}   
%%%%%%%%%%%%%%%%%%%%%%%%%%%%%%
\nn & = -  W_{\alpha}\, W_{\alpha'} \,
\delta(\tau'-\tau'')\,\delta(\mathbf x-\mathbf x ') \sum \limits_{i,j,l}  
 \left( \psi^\dagger_i \, M^i_\alpha \,\psi_i \right )_{\mathbf x,\tau }  
\text{Tr} \left(G_0^j  \, M_{\alpha}^j \, G_0^j \, M^j_{\alpha'} \right )_{\mathbf x,\tau'}
  \left( \psi^\dagger_l \, M^l_{\alpha'} \,\psi_l \right )_{\mathbf x,\tau'''}
%%%%%%%%%%%%%   
\nn & = - n \,
\text{Tr} \left(G_0  \, M_{\alpha} \, G_0 \, M_{\alpha'} \right )_{\mathbf x,\tau'}
W_{\alpha}\, W_{\alpha'} \,
\delta(\tau''-\tau''')  \,\delta(\mathbf x-\mathbf x ')
\sum \limits_{i,j}  \left( \psi^\dagger_i \, M_\alpha^i \,\psi_i \right )_\tau   
\left( \psi^\dagger_j \, M^j_{\alpha'} \,\psi_j \right )_{\mathbf x,\tau'''} .
\end{align}
%%%%%%%%%%%%%%%%%%
Hence, a ZS diagram comes with a factor of $n$, which vanishes upon taking the replica limit $ n \rightarrow 0 $. 
%%%%%%%%%%%%%%%%%%%%%%%%%%%%%%%%%%%%%%%%%%
\item

For the VC diagrams, a sample contraction of terms looks like:
\begin{align}
& W_{\alpha} \, W_{\alpha'} \sum \limits_{i,j,k,l} 
 \left( \psi^\dagger_i \, M_\alpha^i \,\psi_i \right )_{\mathbf x,\tau}   
  \left( \psi^\dagger_j \, M_\alpha^j \,\psi_j \right )_{\mathbf x,\tau'}
\left( \psi^\dagger_k \, M_{\alpha'}^k \,\psi_k \right )_{\mathbf x',\tau''}   
  \left( \psi^\dagger_l \, M_{\alpha'}^l \,\psi_l \right )_{\mathbf x',\tau'''} 
%%%%%%%%%%%%%%%%%%%%%%%%%%%%%%
\nn & = W_{\alpha} \, W_{\alpha'} \,
 \delta(\tau'-\tau'')\,\delta(\tau'-\tau''')\,\delta(\mathbf x-\mathbf x ')
\sum \limits_{i,j} 
 \left( \psi^\dagger_i \, M^i_\alpha \,\psi_j \right )_{\mathbf x, \tau}   
\left( \psi^\dagger_j \, M_{\alpha'}^j \,G_0^j \, M^j_{\alpha} \,G_0^j \, 
M_{\alpha'}^b \,\psi_b \right )_{\mathbf x, \tau''}  .
\end{align}

A VC diagram with two $W_\alpha $ vertices can emerge in 8 distinct way. Gathering all the factors,
and setting the external frequency $p_0 = 0 $ for easily extracting the divergent part,
we find a correction to the 
\\(a) scalar vertex from
\begin{align}
& \Gamma_{00}^{VC}
= 4\,W_0^2\,\mu^{2 \epsilon} \int_{\vec q} G_0(\vec q ,p_0 )\, G_0(\vec q , p_0 )
= \frac{
\left( 1 + \xi ^2 \right) W_0^2  \,\mu^\epsilon}
{2 \,\pi ^2 \left( 1 -\xi ^2 \right)^2 \epsilon }
\left(  \frac{\mu}
{ \sqrt{ \left| p_0\right|}}
\right)^\epsilon .
\end{align}

%%%%%%%%%%%%%%%%%%%%%%%%%%%%%%%%%%%%%%%%%
(b) vector vertex from
\begin{align}
\Gamma_{11}^{VC}
& = 4\,W_1^2\,\mu^{2 \epsilon}\,\Gamma_a^i\,\Gamma_b^j \int_{\vec q} G_0^j(\vec q , p_0 )
\,\Gamma_a^j\,G_0^j(\vec q , p_0 )\,\Gamma_b^j
%%%%%%%%%%%%%%%%%%%%%%%
\nn & = 
-\frac{
\left(N_c-2\right) 
\left(  \frac{2}{N_c} -1 +\xi ^2 \right )
 W_1^2 \,\mu ^{\epsilon}\, \Gamma_a^i\,\Gamma_a^j
}
{2 \,\pi ^2 \left(1-\xi ^2\right)^2  \epsilon}
\left(\frac{\mu }{\sqrt{\left| p_0\right| }}\right)^\epsilon.
\end{align}

(c) tensor vertex from
\begin{align}
\Gamma_{22}^{VC}
& = 4\,W_2^2\,\mu^{2 \epsilon}\,\Gamma_{ab}^i \,\Gamma_{cd}^j \int_{\vec q} G_0^j(\vec q , p_0 )
\,\Gamma_{ab}^j\,G_0^j(\vec q , p_0 )\,\Gamma_{cd}^j
 \nn & 
%%%%%%%%%%%%%%%%%%%%%%%%%%%%%%%%%%%%%%%% 
= \frac{ 
\left(  1 + \xi ^2  -\frac{4}{N_c} \right)
\left(N_c^2-9 N_c+16\right) 
W_2^2 \,\mu ^\epsilon\,\Gamma_{ab}^i\,\Gamma_{ab}^j
}
{4 \,\pi ^2 \left(1-\xi ^2\right)^2  \epsilon }
\left(\frac{\mu }{\sqrt{\left| p_0\right| }}\right)^\epsilon.
\end{align}

VC diagrams with mixed lines can emerge in one of two ways, each of which corrects a different
bare vertex and comes with a combinatorial factor of $4$. The contributions can be listed as:
\\(a) \begin{align}
\Gamma_{01}^{VC}
&= 4\,W_0\,W_1\,\mu^{2 \epsilon} \,\Gamma_b^j\int_{\vec q} G_0^j(\vec q , p_0 )
\, G_0^j(\vec q , p_0 )\,\Gamma_b^j
=
\frac{  \left( 1 + \xi ^2 \right) N_c\, W_0 \,W_1 \,\mu ^{\epsilon}}
{2 \,\pi ^2 \left(1-\xi ^2\right)^2 \epsilon }
\left(\frac{\mu }{\sqrt{\left| p_0\right| }}\right)^\epsilon,
\end{align}
correcting the scalar vertex.

(b)
\begin{align}
\Gamma_{10}^{VC}
= 4\,W_0\,W_1\,\mu^{2 \epsilon}\,\Gamma_a^i\,  \int_{\vec q} G_0^j(\vec q , p_0 )
\,\Gamma_a^j\,G_0^j(\vec q , p_0 ) 
= \frac{ \left( \frac{2}{N_c}- 1+  \xi ^2\right)
\,W_0 \,W_1 \,\mu ^{\epsilon}\, \Gamma_a^i\,\Gamma_a^j}
{2 \,\pi ^2 \left(1-\xi ^2\right)^2 \epsilon}
\left(\frac{\mu }{\sqrt{\left| p_0\right| }}\right)^\epsilon,
\end{align}
correcting the vector vertex.

(c)
\begin{align}
\Gamma_{02}^{VC}
= 4\,W_0\,W_2\,\mu^{2 \epsilon}\,\Gamma_{ab}^j \int_{\vec q} G_0^j(\vec q , p_0 )
\,G_0^j(\vec q , p_0 )\,\Gamma_{ab}^j
=
\frac{\left( 1 +\xi ^2\right) N_c \left(N_c-1\right) 
\,W_0 \,W_2 \,\mu ^{\epsilon }}
{4\, \pi ^2 \left(1-\xi ^2\right)^2 \epsilon }
\left(\frac{\mu }{\sqrt{\left| p_0\right| }}\right)^\epsilon,
\end{align}
correcting the scalar vertex.

(d)
\begin{align}
\Gamma_{20}^{VC}
= 4\,W_0\,W_2\,\mu^{2 \epsilon}\,\Gamma_{ab}^i  \int_{\vec q} G_0^j(\vec q , p_0 )
\,\Gamma_{ab}^j\,G_0^j(\vec q , p_0 ) 
=
\frac{  \left( 1 + \xi ^2  -4/ N_c \right )
W_0 \,W_2 \,\mu^\epsilon\,
\Gamma_{ab}^i\,\Gamma_{ab}^j}
{2 \,\pi ^2 \left(1-\xi ^2\right)^2 \epsilon}
\left(\frac{\mu }{\sqrt{\left| p_0\right| }}\right)^\epsilon,
\end{align}
correcting the tensor vertex.

(e)
\begin{align}
\Gamma_{12}^{VC}
&= 4\,W_1\,W_2\,\mu^{2 \epsilon}\,\Gamma_{a}^i \,\Gamma_{cd}^j \int_{\vec q} G_0^j(\vec q , p_0 )
\,\Gamma_{a}^j\,G_0^j(\vec q , p_0 )\,\Gamma_{cd}^j
\nn &=
\frac{
\left(N_c-4\right) \left(N_c-1\right) 
\left(  \frac{2}{N_c} -1 +\xi^2 \right )
W_1 \,W_2 \,\mu^\epsilon\,\Gamma_{a}^i\,\Gamma_{a}^j
}
{4 \,\pi ^2 \left(1-\xi ^2\right)^2  \epsilon}
\left(\frac{\mu }{\sqrt{\left| p_0\right| }}\right)^\epsilon,
\end{align}
correcting the vector vertex.

(f)
\begin{align}
\Gamma_{21}^{VC}
& = 4\,W_1\,W_2\,\mu^{2 \epsilon}\,\Gamma_{ab}^i \,\Gamma_{c}^j \int_{\vec q} G_0^j(\vec q , p_0 )
\,\Gamma_{ab}^j\,G_0^j(\vec q , p_0 )\,\Gamma_{c}^j
\nn & =
\frac{
\left(  N_c-4\right)
\left(  1 + \xi ^2 -\frac{4}{N_c}\right )
W_1 \,W_2 \,\mu^\epsilon\,
\Gamma_{ab}^i\,\Gamma_{ab}^j}
{2 \,\pi ^2 \left(1-\xi ^2\right)^2 \epsilon}
\left(\frac{\mu }{\sqrt{\left| p_0\right| }}\right)^\epsilon,
\end{align}
correcting the tensor vertex.

%%%%%%%%%%%%%%%%%%%%%%%%%%%%%%%%%%%%%%%%%%%

\item
For the BCS and ZS$^\prime$ diagrams, sample contractions look like:
\begin{align}
&  W_{\alpha} \, W_{\alpha'}  \sum \limits_{i,j,k,l} 
 \left( \psi^\dagger_i \, M_{\alpha}^i \,\psi_i \right )_\tau   
  \left( \psi^\dagger_j \, M_{\alpha'}^j \,\psi_j \right )_{\tau'}
%%%
\left( \psi^\dagger_k \, M_{\alpha}^k \,\psi_k \right )_{\tau''}   
  \left( \psi^\dagger_l \, M_{\alpha'}^l \,\psi_l \right )_{\tau'''}  
%%%%%%%%%%%%%%%%%%%%%%%%%%%%%%
\nn & = W_{\alpha} \, W_{\alpha'} \,\delta(\tau-\tau')\, \delta(\tau''-\tau''')\sum \limits_{i,j}
\left( \psi^\dagger_i \, M_{\alpha}^i \,G_0^i \, M_{\alpha'}^i \,\psi_i \right )_{\tau}
\left( \psi^\dagger_j \, M_{\alpha}^j \,G_0^j \, 
M_{\alpha'}^j \,\psi_j \right )_{\tau''} ,
\end{align}
%%%%%%%%%%%%%%5555
and
\begin{align}
W_{\alpha} \, W_{\alpha'} \,
\delta(\tau-\tau')\, \delta(\tau''-\tau''')\sum \limits_{i,j}
\left( \psi^\dagger_i \, M_{\alpha}^i \,G_0^i \, M_{\alpha'}^i \,\psi_a \right )_{\tau}
\left( \psi^\dagger_j \, M_{\alpha'}^j \,G_0^j \, 
M_{\alpha}^j \,\psi_j \right )_{\tau''} ,
\end{align}
%%%%%%%%%%%%%%%%%%%%%%%
respectively.

Each of these one-loop corrections comes with an overall factor of $2$, and the final answer depends on the nature of the internal disorder lines. The contributions, combining the BCS and ZS' counterparts, can be listed as:
\\(a)
\begin{align}
\Pi_{00}^{BCS+ZS'}
= 4\,W_0^2\,\mu^{2 \epsilon} \int_{\vec q} G_0^i(\vec q , p_0 ) \, G_0^j (\vec q , p_0 )
= \left[
\frac{\xi ^2}
{2\, \pi ^2 \left(1-\xi ^2\right)^2 \epsilon }
+
\frac{\Gamma _a^i \,\Gamma _a^j}
{2 \,\pi ^2 \left(1-\xi ^2\right)^2 N_c\,\epsilon  }
\right]
W_0^2 \,\mu ^{\epsilon }
\left(\frac{\mu }{\sqrt{\left| p_0\right| }}\right)^\epsilon,
\end{align}
correcting both the scalar and vector vertices.

(b)
\begin{align}
\Pi_{11}^{BCS+ZS'}
& =2\,W_1^2\,\mu^{2 \epsilon} \int_{\vec q} 
 \Gamma_{a}^i \,G_0^i (\vec q , p_0 )\, \Gamma_{b}^i 
\left[  \Gamma_{a}^j(\vec q , p_0 ) \,G_0^j(\vec q , p_0 ) \, \Gamma_{b}^j 
+  \Gamma_{b}^j \,G_0^j(\vec q , p_0 ) \, \Gamma_{a}^j \right]
\nn & = \left[
\frac{\xi ^2 \,N_c}
{2 \,\pi ^2 \left(1-\xi ^2\right)^2 \epsilon }
+
\frac{\left(3 \,N_c-2\right) \Gamma _a^i \,\Gamma _a^j}
{2\, \pi ^2 \left(1-\xi ^2\right)^2 N_c\,\epsilon }
\right]
W_1^2 \,\mu ^{\epsilon }
\left(\frac{\mu }{\sqrt{\left| p_0\right| }}\right)^\epsilon,
\end{align}
correcting both the scalar and vector vertices.

(c)
\begin{align}
\Pi_{22}^{BCS+ZS'}
&= 2\,W_2^2\,\mu^{2 \epsilon} \int_{\vec q} 
 \Gamma_{ab}^i \,G_0^i (\vec q , p_0 )\, \Gamma_{cd}^i 
 \left[ \Gamma_{ab}^j \,G_0^j (\vec q , p_0 )\, \Gamma_{cd}^j 
+ \Gamma_{cd}^j \,G_0^j (\vec q , p_0 )\, \Gamma_{ab}^j \right]
\nn & =
\left[
\frac{ 6 + 13 \,\xi ^2}
{2 \,\pi ^2 \left(1-\xi ^2\right)^2 \epsilon }
+
\frac{\left(17 + 15\, \xi ^2\right) \Gamma _a^i \,\Gamma _a^j}
{5 \,\pi ^2 \left(1-\xi ^2\right)^2 \epsilon }
-\frac{3 \,\xi ^2\,\Gamma_{ab}^i \,\Gamma _{ab}^j }
{\pi ^2 \left(1-\xi ^2\right)^2 \epsilon }
\right]
W_2^2 \,\mu ^{\epsilon }
\left(\frac{\mu }{\sqrt{\left| p_0\right| }}\right)^\epsilon,
\end{align}
correcting the scalar, vector, and tensor vertices.

(d)
\begin{align}
\Pi_{01}^{BCS+ZS'}+\Pi_{10}^{BCS+ZS'}
&= 4\,W_0\,W_1\,\mu^{2 \epsilon} \int_{\vec q} G_0^i (\vec q , p_0 )\, \Gamma_{a}^i 
\left[  G_0^j (\vec q , p_0 )\, \Gamma_{a}^j +  \Gamma_{a}^j \,G_0^j (\vec q , p_0 ) \right]
%%%%%5
\nn & =
\left[
\frac{\xi ^2}
{\pi ^2 \left(1-\xi ^2\right)^2 \epsilon }
+
\frac{\Gamma _a^i \,\Gamma _a^j}
{\pi ^2 \left(1-\xi ^2\right)^2 N_c\, \epsilon }
\right]
W_0 \,W_1 \,\mu ^{\epsilon }
\left(\frac{\mu }{\sqrt{\left| p_0\right| }}\right)^\epsilon,
\end{align}
correcting both the scalar and vector vertices.

(e)
\begin{align}
\Pi_{02}^{BCS+ZS'}+\Pi_{20}^{BCS+ZS'}
& = 4 \,W_0\,W_2\,\mu^{2 \epsilon} \int_{\vec q} G_0^i (\vec q , p_0 )\, \Gamma_{ab}^i 
\left[  G_0^j(\vec q , p_0 ) \, \Gamma_{ab}^j +  \Gamma_{ab}^j \,G_0^j (\vec q , p_0 ) \right]
%%%%%5
\nn & =
\frac{\left(  3 /10 + \xi ^2\right) \Gamma _{\text{ab}}^i \,\Gamma _{\text{ab}}^j}
{\pi ^2 \left(1-\xi ^2\right)^2 \epsilon }
\, W_0 \,W_2 \,\mu ^{\epsilon }
\left(\frac{\mu }{\sqrt{\left| p_0\right| }}\right)^\epsilon,
\end{align}
correcting the tensor vertex.

(f)
\begin{align}
\Pi_{12}^{BCS+ZS'} + \Pi_{21}^{BCS+ZS'}
& = 4\,W_1\,W_2\,\mu^{2 \epsilon} \int_{\vec q} 
 \Gamma_{a}^i \,G_0^i(\vec q , p_0 ) \, \Gamma_{bc}^i 
\left[  \Gamma_{a}^j \,G_0^j (\vec q , p_0 )\, \Gamma_{bc}^j 
+  \Gamma_{bc}^j \,G_0^j(\vec q , p_0 ) \, \Gamma_{a}^j \right]
%%%%%5
\nn & =
\frac{  34 + 15\, \xi ^2 }
{10 \,\pi ^2 \left(1-\xi ^2\right)^2 \epsilon }\,
W_1 \,W_2 \,\mu ^{\epsilon }
\left(\frac{\mu }{\sqrt{\left| p_0\right| }}\right)^\epsilon,
\end{align}
correcting the tensor vertex.

\end{enumerate}

For calculational convenience, the final results have been gathered and tabulated in Tables~\ref{tabvc} and \ref{tabBCS}.

%%%%%%%%%%%%%%%%%%TABLE%%%%%%%%%%%%%%%%%%%%%%%%
\begin{table}
\begin{tabular}{|c|c|c|c|}
\hline
Coupling  & $ W_0$ & $ W_1 $ & $ W_2$  \tabularnewline
\hline
$ W_0$ & 
$ Z_{7,1} \rightarrow
 \frac{ - \left( 1  + \xi^2 \right)  W_0 } 
{2\,\pi^2 \left( 1 - \xi^2 \right)^ 2} $
& 
$ Z_{7,1} \rightarrow
- \frac{  \left( 1 + \xi ^2 \right) N_c\, W_1}
{2 \,\pi ^2 \left(1-\xi ^2\right)^2  }
$
&$ Z_{7,1} \rightarrow
-
\frac{\left( 1 + \xi ^2\right) N_c \left(N_c-1\right) W_2}
{4 \,\pi ^2 \left(1-\xi ^2\right)^2  }
$
 \tabularnewline
\hline
$ W_1$ &
$  Z_{8,1} \rightarrow
 \frac{ \left (
\frac{2}{N_c}-1 + \xi ^2 \right ) W_0}
{2 \,\pi ^2 \left(1-\xi ^2\right)^2}
$
&
$  Z_{8,1} \rightarrow
 \frac{
\left(N_c-2\right) 
\left(  \frac{2}{N_c} -1 +\xi ^2 \right )
 W_1 }
{2 \,\pi ^2 \left(1-\xi ^2\right)^2} $
& 
$  Z_{8,1} \rightarrow
- 
\frac{ 
\left(N_c-4\right) \left(N_c-1\right) 
\left (  \frac{2}{N_c} -1 +\xi^2   \right ) W_2 }
{4 \,\pi ^2 \left(1-\xi ^2\right)^2  }
$
 \tabularnewline
\hline
$ W_2$ &  
 $ Z_{9,1}  \rightarrow
- \frac{   \left(  1  + \xi ^2 -\frac{4}{N_c}
\right ) W_0 }
{2 \,\pi ^2 \left(1-\xi ^2\right)^2  }
$
&
$ Z_{9,1}  \rightarrow
-
\frac{ \left(N_c-4\right)\left( 1  + \xi ^2 -\frac{4}{N_c} \right ) W_1 }
{2 \,\pi ^2 \left(1-\xi ^2\right)^2 \epsilon }
$
&
$  Z_{9,1} \rightarrow - \frac{ 
\left(  1 + \xi ^2  -\frac{4}{N_c} \right)
\left(N_c^2-9 N_c+16\right) W_2 }
{4 \,\pi ^2 \left(1-\xi ^2\right)^2  }
$
 \tabularnewline
\hline
\end{tabular}
\caption{\label{tabvc}Contributions to $Z_{n,1}$ (for $n=7,\,8, \,9$) from the disorder-only VC diagrams.}
\end{table}

%%%%%%%%%%%%%%%%%%%%%%%%%%%%%%
 \begin{table}
\begin{tabular}{|c|c|c|c|}
\hline
Coupling  & $ W_0$ & $ W_1 $ & $ W_2 $  \tabularnewline
\hline
$ W_0$ &
\makecell{
$Z_{7,1}  \rightarrow
- \frac{\xi ^2 \,W_0}
{2\, \pi ^2 \left(1-\xi ^2\right)^2  }
$ \\
$Z_{8,1}  \rightarrow
- 
\frac{W_0^2 }
{2 \,\pi ^2 \left(1-\xi ^2\right)^2  N_c \,W_1}
$
}
%%%%%%%%%%%%%
& 
\makecell{
$Z_{7,1}  \rightarrow
- 
\frac{\xi ^2 \, W_1}
{\pi ^2 \left(1-\xi ^2\right)^2 }
$ \\
$Z_{8,1}  \rightarrow
- 
\frac{W_0 }
{\pi ^2 \left(1-\xi ^2\right)^2  N_c}
$}
%%%%%%%%%%%%%%%%%
&
$Z_{9,1}  \rightarrow
-
\frac{\left(\xi ^2+\frac{3}{10}\right) W_0  }
{\pi ^2 \left(1-\xi ^2\right)^2 }
$
 \tabularnewline
\hline
$W_1$ &
\makecell{included in $( W_0, \, W_1)$ cell}
%%%%%%%%%%%%%%%%%
& \makecell{
$Z_{7,1}  \rightarrow -
\frac{\xi ^2 \,N_c\, W_1^2}
{2 \,\pi ^2 \left(1-\xi ^2\right)^2 W_0 }
$ \\
$Z_{8,1}  \rightarrow -
\frac{ \left(3 \,N_c-2\right) W_2 }
{2 \,\pi ^2 \left(1-\xi ^2\right)^2   N_c}
$
}
%%%%%%%%%%%%%%%%%%%%%%%55
& 
$Z_{9,1}  \rightarrow -
\frac{  \left(34  + 15\, \xi ^2 \right) W_1
}
{10 \,\pi ^2 \left(1-\xi ^2\right)^2  }
$
\tabularnewline
\hline
%%%
$W_2 $ & \makecell{included in $( W_0, \, W_2)$ cell}
& \makecell{included in $( W_1, \,W_2)$ cell} &
%%%%%%%%%%%%%%%%%%%
\makecell{
$Z_{7,1}  \rightarrow -
\frac{ \left( 6 + 13 \,\xi ^2 \right) W_2^2}
{2 \,\pi ^2 \left(1-\xi ^2\right)^2 W_0 } $\\
%%%%%%%
$Z_{8,1}  \rightarrow -
\frac{  \left( 17  + 15 \,\xi ^2 \right)  W_2^2 }
{5 \,\pi ^2 \left(1-\xi ^2\right)^2 W_1 }
$ \\
%%%%%%%%%%%%%%%
$Z_{9,1}  \rightarrow 
\frac{3 \,\xi ^2\,W_2}
{\pi ^2 \left(1-\xi ^2\right)^2  }
$
}
%%%%%%%%%%%%%%%%%%%%%%%%
\tabularnewline
\hline
\end{tabular}
\caption{\label{tabBCS}Contributions to $Z_{n,1}$ (for $n=7,\,8, \,9$)
from the BCS and ZS$'$ diagrams.}
\end{table} 
%%%%%%%%%%%%%%%%%%%%%%%%%%%%%%%

%%%%%%%%%%%%%%%%%%%%%%%%
\subsection{RG equations}

The counterterm action for the disorder part is given by:
\begin{align}
 \label{disac} 
 {\mathcal S}^{\text{dis}}_{CT}= 
& - \mu^ \epsilon \sum \limits_{i,j }   \int_{p_0} \int_{k_0}
\left( \prod \limits_{ m=1}^4 \int_{\vec{p}_m } \right) \left(2\,\pi\right)^{d} \,\delta^d\left( \vec p_1 + \vec p_3-\vec p_2-\vec p_4 \right)
\nn & \hspace{ 1.5 cm } \times \Big [ \,
A_7 \, W_0   \left\lbrace \psi_i^{\dag} (\vec p_1 , p_0) \, {\psi_i} (\vec p_2 , p_0) \right \rbrace
\left\lbrace \psi_j^{\dag} (\vec p_3 , k_0) \, {\psi_j } (\vec p_4 , k_0) \right \rbrace
\nn &  \hspace{ 2 cm }  +   A_8 \,W_1 \sum \limits_{a} 
    \left\lbrace \psi_i^{\dag} (\vec p_1 , p_0) \,\Gamma_a^i\, {\psi_i} (\vec p_2 , p_0) \right \rbrace  
\left\lbrace \psi_j^{\dag} (\vec p_3 , k_0) \,\Gamma_a^j\, {\psi_j } (\vec p_4 , k_0) \right \rbrace
\nn
 &   \hspace{ 2 cm } +  A_9\,   W_2 \sum \limits_{a<b}
\left\lbrace \psi_i^{\dag} (\vec p_1 , p_0) \,\Gamma_{ab}^i  \, {\psi_i} (\vec p_2 , p_0) \right \rbrace
\left\lbrace \psi_j^{\dag} (\vec p_3 , k_0) \,\Gamma_{ab}^j  \, {\psi_j } (\vec p_4 , k_0) \right \rbrace \,\Big ] \,,
%%%%%%%%%%%%
\end{align}
where $A_n= Z_n-1  =  \sum \limits_{\lambda=1}^\infty
\frac{Z_{n,\lambda}} {\epsilon^\lambda} $ as before. 
%%%%%%%%%%%%%%%%%%%%%%%%%%%%%%%%%%

Adding these counterterms, the disorder part of the renormalized action is:
\begin{align}
 \label{dis-ren} 
{\mathcal S}^{\text{dis}}_{ ren }= 
&  - \sum \limits_{i,j }   \int_{p_{0_B}} \int_{k_{0_B}}
\left( \prod \limits_{ m=1}^4 \int_{\vec{p}_{m_B} } \right) \left(2\,\pi\right)^{d} \,\delta^d\left( \vec p_{1_B} + \vec p_{3_B}-\vec p_{2_B}-\vec p_{4_B} \right)
\nn & \hspace{ 1.0 cm } \times \Big [ \,
 {W_0}_B  \left\lbrace {\psi_i}_B^{\dag} (\vec p_{1_B} , p_{0_B} ) \, {\psi_i}_B (\vec p_{2_B} , p_{0_B}) \right \rbrace
\left\lbrace  {\psi_j}_B^{\dag} (\vec p_{3_B} , k_{0_B} ) \,  {\psi_j}_B (\vec p_{4_B} , k_{0_B} ) \right \rbrace
\nn &  \hspace{ 1.5 cm }  
+   {W_1}_B \sum \limits_{a} 
    \left\lbrace {\psi_i}_B^{\dag} (\vec p_{1_B} , p_{0_B} ) \,\Gamma_a^i\, {\psi_i}_B(\vec p_{2_B} , p_{0_B} ) \right \rbrace
\left\lbrace  {\psi_j}_B^{\dag} (\vec p_{3_B} , k_{0_B} ) \,\Gamma_a^j\,  {\psi_j}_B(\vec p_{4_B} , k_{0_B} ) \right \rbrace
\nn
 &   \hspace{ 1.5 cm } +    {W_2}_B \sum \limits_{a<b}
\left\lbrace{\psi_i}_B^{\dag} (\vec p_{1_B} , p_{0_B} )   \,\Gamma_{ab}^i  \, {\psi_i}_B(\vec p_{2_B} , p_{0_B} ) \right \rbrace
\left\lbrace  {\psi_j}_B^{\dag} (\vec p_{3_B} , k_{0_B} )  \,\Gamma_{ab}^j  \, {\psi_j}_B (\vec p_{4_B} , k_{0_B} ) \right \rbrace \Big ] \,,
%%%%%%%%%%%%
\end{align}
%%%%%%%%%%%%%%%%%%%%%%%%%%%%%%%%%%
where, in addition to Eq.~(\ref{scale1}), we now have:
\begin{align}
{W_0}_B= \frac{Z_7 \,\mu ^{\epsilon }}
{Z_2^{d/2}}  \, W_0 \,,
\quad {W_1}_B= \frac{Z_8 \,\mu ^{\epsilon }}
{Z_2^{d/2}} 
 \, W_1\,,
\quad {W_2}_B=
\frac{Z_9 \,\mu ^{\epsilon }}
{Z_2^{d/2}}  \, W_2\,.
\end{align}

The disorder beta functions are now obtained from $\frac{d \ln {W_\alpha}_B}{d \ln \mu}=0$, which take the form:
\begin{align}
-\frac{\partial W_0}{\partial \ln \mu}\equiv -\beta_{W_0} & =
\Big [  \epsilon  + \frac{\partial \ln Z_7 }{\partial \ln \mu} 
+d \left(1-z_r\right)
\Big ] \,W_0 \,,\nn
%%%%%%%%%%%%%%%%
-\frac{\partial W_1}{\partial \ln \mu}\equiv -\beta_{W_0} & =
\Big [  \epsilon  + \frac{\partial \ln Z_8  }{\partial \ln \mu} 
+d \left(1-z_r\right)
\Big ] \,W_1 \,,\nn
-\frac{\partial W_2 }{\partial \ln \mu}\equiv -\beta_{W_0} & =
\Big [  \epsilon  + \frac{\partial \ln Z_9 }{\partial \ln \mu} 
+d \left(1-z_r\right)
\Big ] \,W_2 \,.
\end{align}

Using all the loop-calculation results (see also Tables~\ref{tabvc} and \ref{tabBCS}), we find that:
\begin{align}
 \label{rgmy2} 
Z_{1,1} &= - \frac{g^2 \,y }
{ 16 \,\pi ^2 \left ( 1 +y -\xi\, y \right )^2} 
+ 
\frac{\left( 1+\xi^2 \right)  \left [ W_0  + N_c\, W_1 + \frac{N_c \left(N_c-1\right)}{2}\right ]}
{4\,\pi^2  \left(1-\xi^2 \right)^2} 
\,,\quad
%%%%%%%%%%
\nn Z_{2,1}  & =  \frac{g^2}
 { 48 \,\pi ^2 \left ( 1 +y -\xi\, y \right )^3}  \,,\quad
%%%%%%%%%%%
Z_{3,1}  = - \frac{g^2 \,y  \left (1-\xi \right ) }
{  16 \,\pi ^2 \left ( 1 +y -\xi\, y \right )^3 \xi}  \,,\nn
 %%%%%%%%%
Z_{4,1}  &=   \frac{13 \,N\,g^2 \,\xi }
{288 \, \pi ^2 \left(1-\xi ^2\right)^2 y} \,,\quad
%%%%%%%%%%%%%%%%5
Z_{5,1} = -\frac{107 \,N\, g^2 }
{3456 \,\pi ^2 \left(1-\xi ^2\right)} \,,   \quad
 Z_{6,1} = 
 -\frac{ \left( 1+\xi^2 \right)  
\left [  W_0 +  N_c \, W_1  -\frac{N_c \left( N_c-1 \right) W_2 }{2} \right ]}
{ 4\,\pi^2 \left( 1-\xi^2 \right)}  \,,\nn
%%%%%%%%%%%%%%%%%%%%%%%%%%%%%%%%%%%%%%%%%%%%
Z_{7,1}   & =  -\frac{g^2 \,y} {4\, \pi ^2 \left ( 1 +y -\xi\, y \right )^2}
\nn & \qquad -\frac{\xi ^2 \left(W_1^2 \,N_c
+ 2 \,W_0^2 + 7\, W_1 \,W_0
+10 \,W_2\, W_0
+13 \,W_2^2\right)
+ W_0\, W_1 \,N_c
+W_0^2
+6 \,W_2^2 +10 \,W_0 \,W_2}
{2 \,\pi ^2 \left( 1 -\xi ^2 \right)^2 W_0} \,,\nn
%%%%%%%%%%
Z_{8,1}  &=
-\frac{g^2 \left[
1-\xi \,y \left (N_c-1 \right ) -N_c \left (1 + y \right ) + 2\, y
\right ]}
{4\, \pi ^2\, N_c \left ( 1 +y -\xi\, y \right )^2}
%%%
\nn & \qquad -\frac{  N_c\, \xi ^2 
\left [ 15\, W_2^2+2 \,W_1\, W_2
+W_1 \left(W_0-3\, W_1\right)\right ]
+W_0^2+9 \,W_1^2+85\, W_2^2-W_0 \,W_1+7\, W_1 \,W_2}
{10\, \pi ^2 \left( 1 - \xi ^2 \right)^2 W_1}
 \,,\nn
%%%%%%%%%%%
\nn Z_{9,1} & = 
-\frac{g^2 \left[
2-y \left (2 \,\xi +N_c-4 \right ) \right ]}
{4\, \pi ^2 \,N_c \left ( 1 +y -\xi\, y \right )^2}
%%%
-\frac{ N_c \,\xi ^2 \left(3 \,W_0+4\, W_1-8 \,W_2\right)
+4 \,W_0+35 \,W_1-2 \,W_2}
{10\, \pi ^2 \left( 1 -\xi ^2 \right)^2}  \,.
\end{align}

Finally, using the expansions in Eq.~(\ref{expand0}), and
\begin{align}
&
\beta_{W_0}=\beta_{W_0}^{(0)} +\epsilon \, \beta_{W_0}^{(1)}\,,
\quad \beta_{W_1}=\beta_{W_1}^{(0)} +\epsilon \, \beta_{W_1}^{(1)}\,, 
\quad \beta_{W_2}=\beta_{W_2}^{(0)} +\epsilon \, \beta_{W_2}^{(1)}\,,
\label{expand}
\end{align}
we get:
\begin{align}
&  z_\tau =2 
+ \frac{g^2\, y} {16 \,\pi ^2 \left (  1 +y -\xi\, y \right )^2}
-\frac{\left( 1 + \xi ^2\right) 
\left[ N_c \left(W_1+2\, W_2\right)  + W_0\right] }
{4 \,\pi ^2 \left( 1-\xi ^2\right)^2}
\,,\quad
z_r = 1 -\frac{g^2} {96 \,\pi ^2 \left (  1 +y -\xi\, y \right )^3}
\,,\nn
%%%%%%%%
& \eta_\psi = - \frac{g^2\,\epsilon} {192 \, \pi ^2 \left (  1 +y -\xi\, y \right )^3}
+
\frac{g^2} {48\, \pi ^2 \left (  1 +y -\xi\, y \right )^3}
-\frac{g^2 \,y} {32 \,\pi ^2 \left (  1 +y -\xi\, y \right )^2}
+
\frac{\left( 1 +\xi ^2\right)
\left[ N_c \left(W_1+2 \,W_2\right )+W_0 \right ]}
{8\, \pi ^2 \left(1-\xi ^2\right)^2}
\,,\nn
%%%%%%%%%%%%%
& \eta_\phi = -\frac{g^2\,\epsilon} {192 \, \pi ^2 \left (  1 +y -\xi\, y \right )^3}
+
\frac{g^2}{\pi ^2} \left[ 
\frac{107 \,N} {6912 \left(1-\xi ^2\right)}
-\frac{y} {32 \left (  1 +y -\xi\, y \right )^2}
+\frac{1} {32 \left (  1 +y -\xi\, y \right )^3}
\right ]
%%%
+
\frac{\left( 1+\xi ^2 \right) 
\left[ N_c \left(W_1+2 \,W_2\right) + W_0\right ]}
{8 \,\pi ^2 \left( 1 -\xi ^2 \right)^2}\,,
\end{align}
and
the beta-functions:
%%%%%%%%%%%%%%%%%%
\begin{align}
 \beta_\xi & = 
-\frac{\xi +3 \left (1-\xi \right ) y} 
{48\, \pi ^2 \left ( 1 +y -\xi\, y \right )^3}\,g^2 \,, \nn 
%%%%%%%%%%%%%%%%%%%%%%%%
  \beta_y & =  \frac{g^2} {\pi ^2}
\left[
\frac{13\,N \,\xi }
{288 \left(1-\xi ^2\right)^2}
+\frac{107 \,N\, y}
{3456 \left(1-\xi ^2\right)}
+\frac{y^2} {16 \left ( 1 +y -\xi\, y \right )^2}
+\frac{y} {48 \left ( 1 +y -\xi\, y \right )^3}
\right ] 
%%%
-\frac{ y \left( 1 + \xi ^2 \right) 
 \left[  N_c \left(W_1+2 \, W_2\right)+W_0\right ]}
{4 \,\pi ^2 \left( 1 -\xi ^2 \right)^2}
 \,,\nn 
%%%%%%%%%%%%%%%%%%%%%%%%%%%%%%%%%
\beta_g & =- \frac{\epsilon \,g }  {2} 
+ 
\frac{g^3} {\pi ^2}
 \left[ 
 \frac{107\, N } {6912 \left(1-\xi ^2\right)}
 +\frac{y} {32 \left ( 1 +y -\xi\, y \right )^2}
 -\frac{1} {96 \left ( 1 +y -\xi\, y \right )^3} \right ]
 %%%%%
 -\frac{g  \left(3 \,W_0+15\, W_1-10\, W_2\right)
\left( 1+\xi ^2 \right) }
 {8 \,\pi ^2 \left(1-\xi ^2\right)^2}  \,,\nn
%%%%%%%%%%%%%%%%%%%%%%%%%%%
%%%%%%%%%%%%%%%%%%%
 \beta_{W_0} & = - \epsilon \,W_0
-\frac{g^2\,W_0} {4\,\pi^2 \left (1 +y -\xi\, y \right )^2}
 \left[
\frac{1}{6 \left (1 +y -\xi\, y \right )} -y
 \right ]
%%%
\nn & \qquad -\frac{\xi ^2 
\left[ N_c\, W_1^2 +2 \,W_0^2+7 \,W_1\, W_0+10 \,W_2\, W_0 + 13\, W_2^2\right ]
+  N_c  \,W_0 \,W_1+W_0^2+6 \,W_2^2 + 10\,W_0\, W_2}
{2 \,\pi ^2 \left(1-\xi ^2\right)^2} 
  \,,\nn
%%%%%%%%%%%%%%%%%%%%%%%%%%%%%%%%%%
\beta_{W_1} & = - \epsilon \,W_1 
+g^2\, W_1\,
\frac{ 29 + 6 \,y \left[ 7 + y \left(3-4 \,\xi^2+\xi \right)
\right ]}
{120 \,\pi ^2 \left (1 +y -\xi\, y \right )^3}
%%%
\nn & \qquad -\frac{\xi ^2 
\left[15 \,W_2^2 + 2 \,W_1\, W_2+ N_c \,W_1 \left(W_0-3\, W_1\right)\right ]
+W_0^2+9 \,W_1^2+85 \,W_2^2-W_0 \,W_1+7\, W_1\, W_2
}
{10 \,\pi ^2 \left( 1 -\xi ^2\right)^2}
\,,\nn
%%%%%%%%%%%%%%%%%%%
\beta_{W_2 } & = - \epsilon \,W_2 
- W_2 \left[
g^2\,
\frac{  7 + 6 \,y \left \lbrace 1 -4 \,\xi + \left (\xi -1 \right ) 
\left ( 1 + 2 \,\xi  \right ) y \right \rbrace 
}
{120 \, \pi ^2 \left (1 +y -\xi\, y \right )^3}
+\frac{\xi ^2 \left(3 \,W_0+4 \,W_1-8 \,W_2\right) N_c+4 \,W_0+35\, W_1-2 \,W_2}
{10\, \pi ^2 \left( 1 -\xi ^2\right)^2}
\right ]\,.
\label{betafinal}
\end{align}

%%%%%%%%%%%%%%%%%%%%%%%%%%%%%%%%%%%%%%%%%
\begin{figure}[]
\centering
\subfigure[]{\includegraphics[width = 0.33\columnwidth]{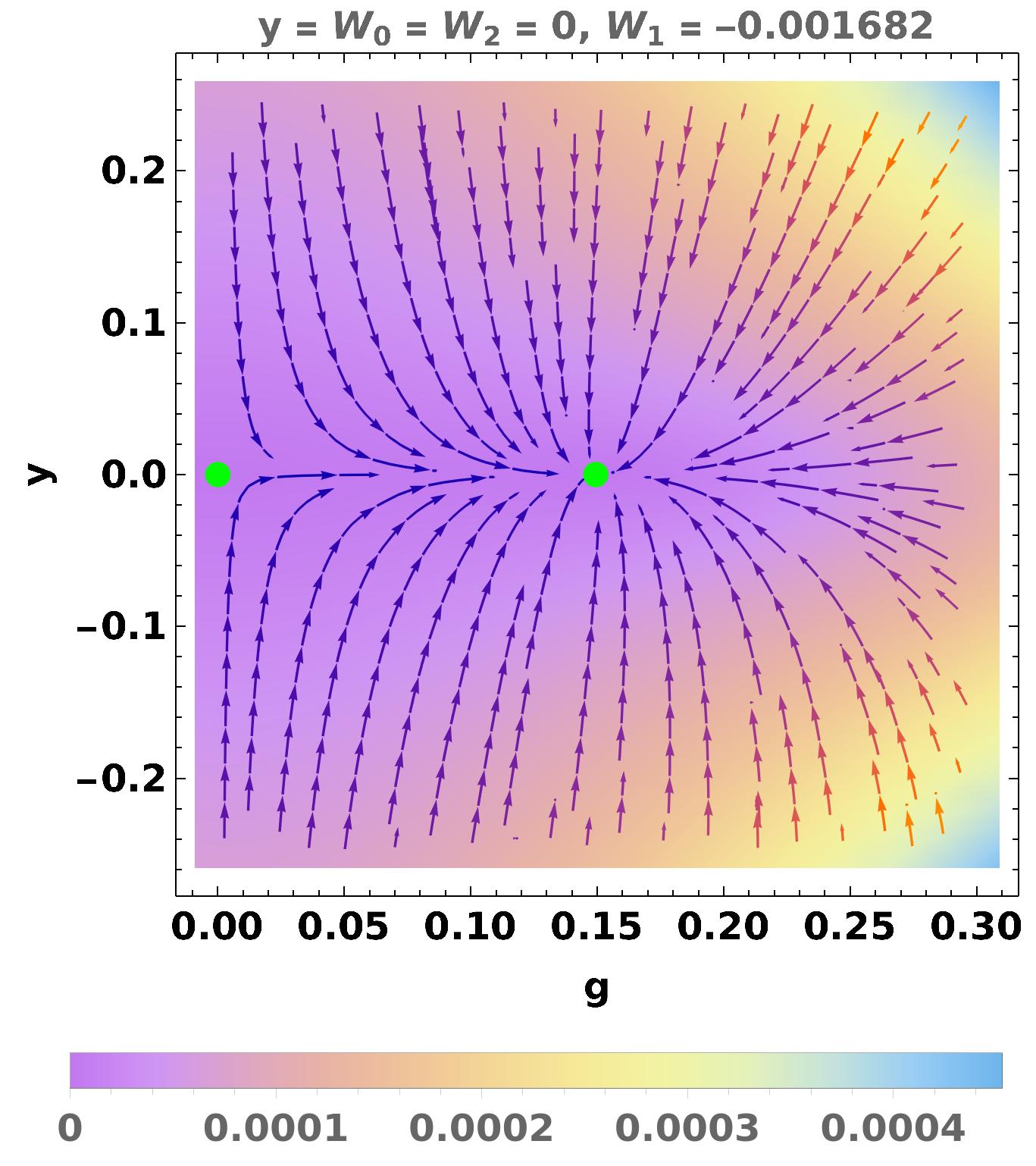}} \hspace{1 cm}
\subfigure[]{\includegraphics[width = 0.339 \columnwidth]{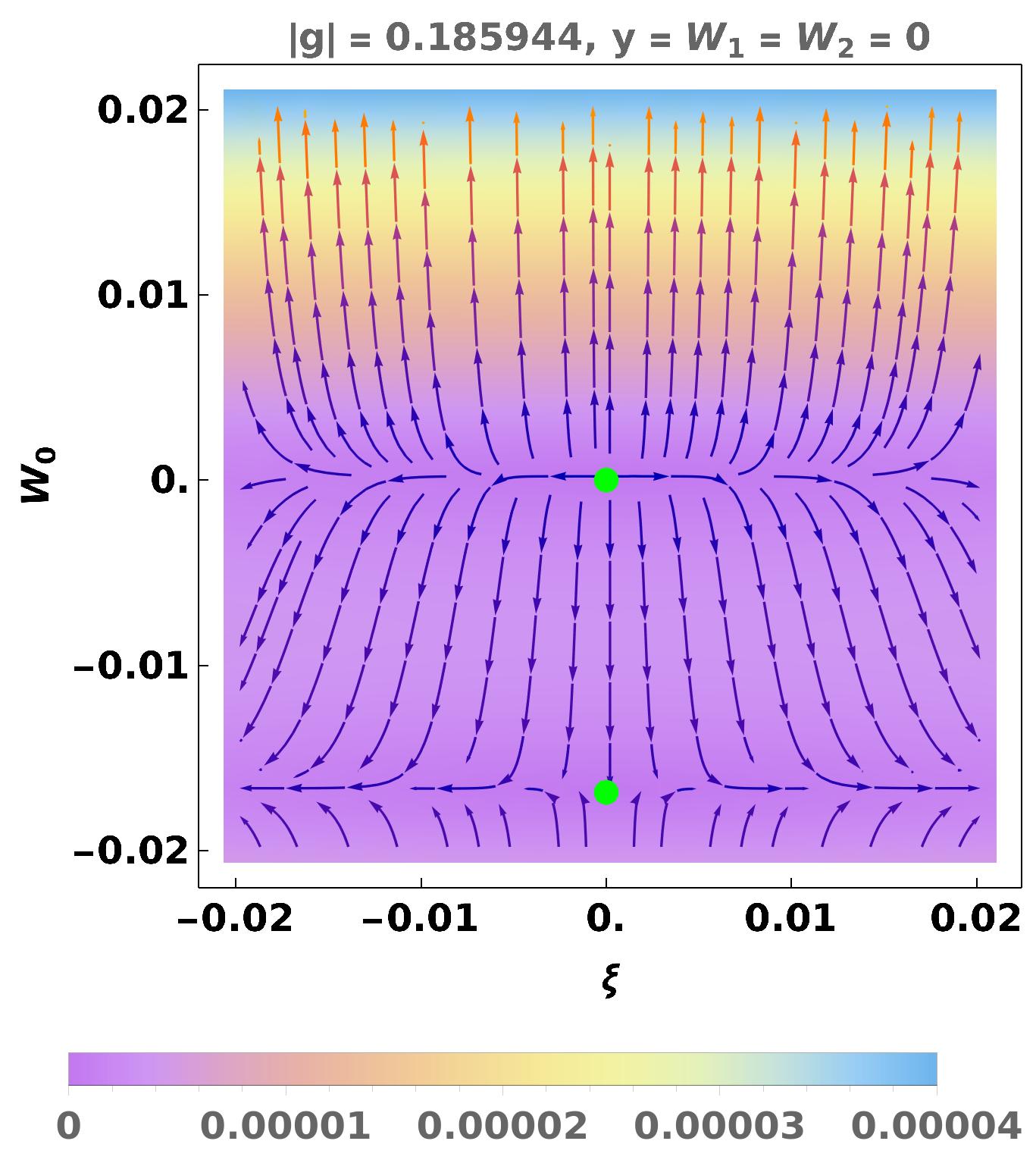}} \\
\subfigure[]{\includegraphics[width = 0.343 \columnwidth]{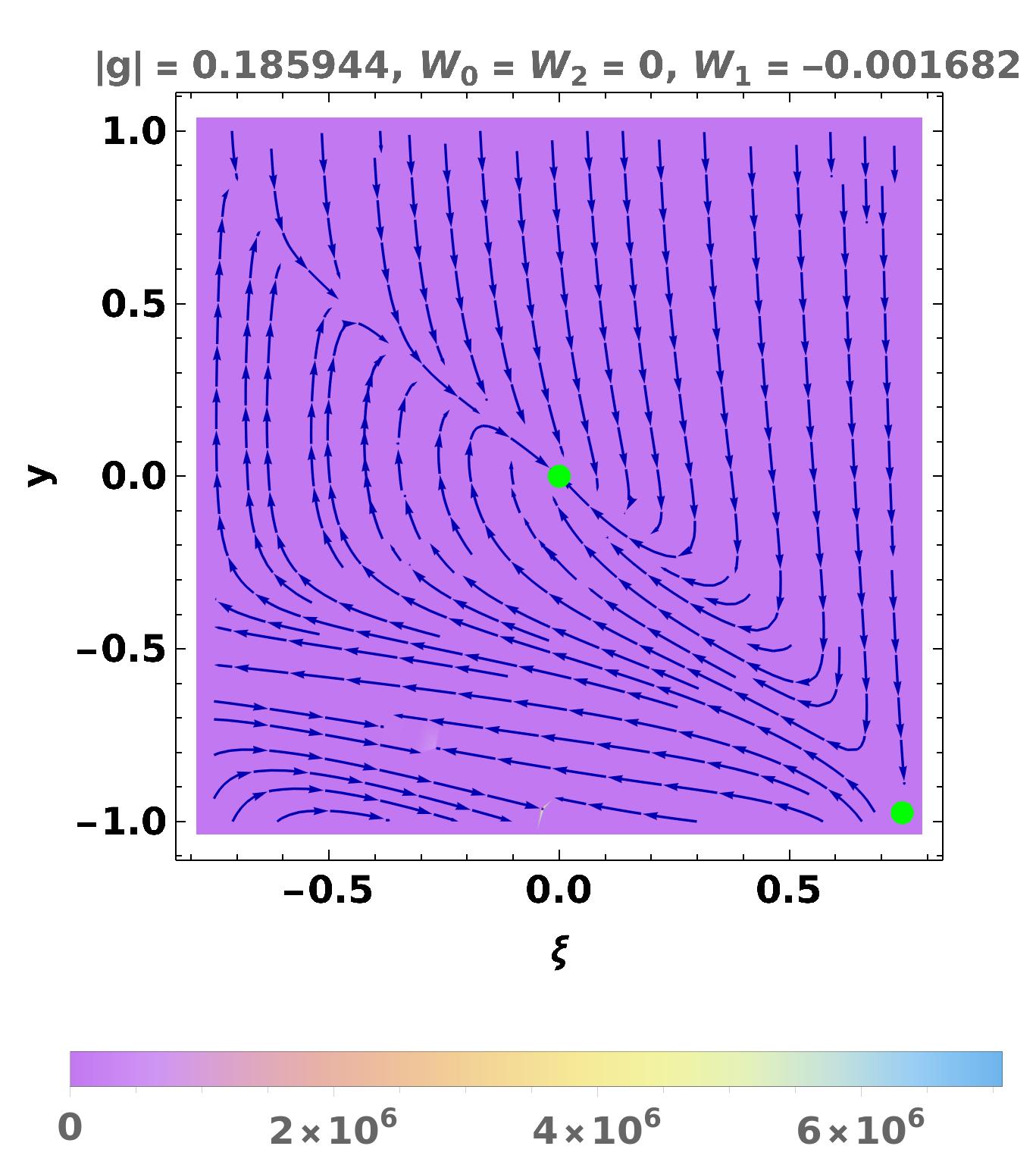}} \hspace{1 cm}
\subfigure[]{\includegraphics[width = 0.33 \columnwidth]{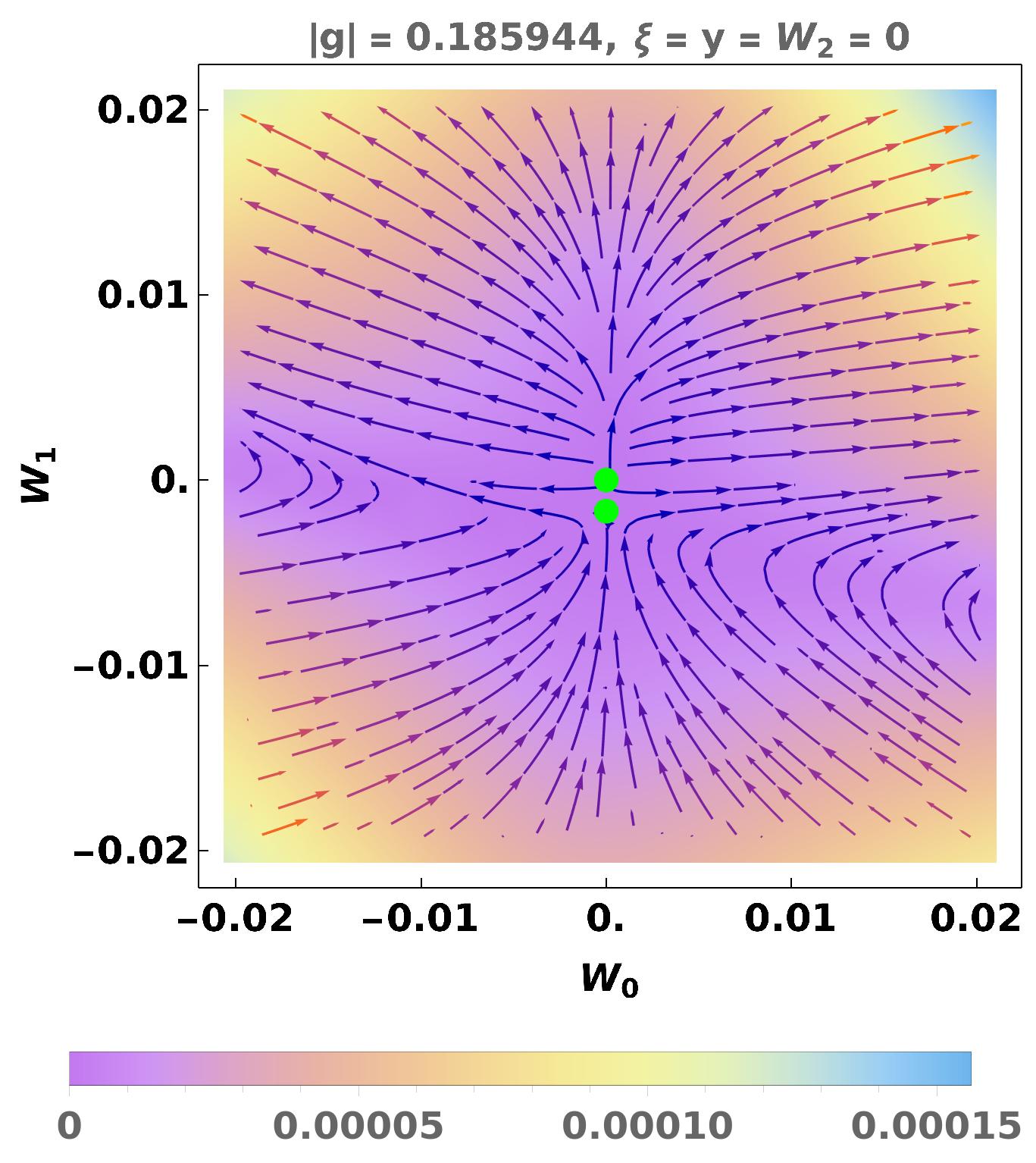}}\\
\subfigure[]{\includegraphics[width = 0.335 \columnwidth]{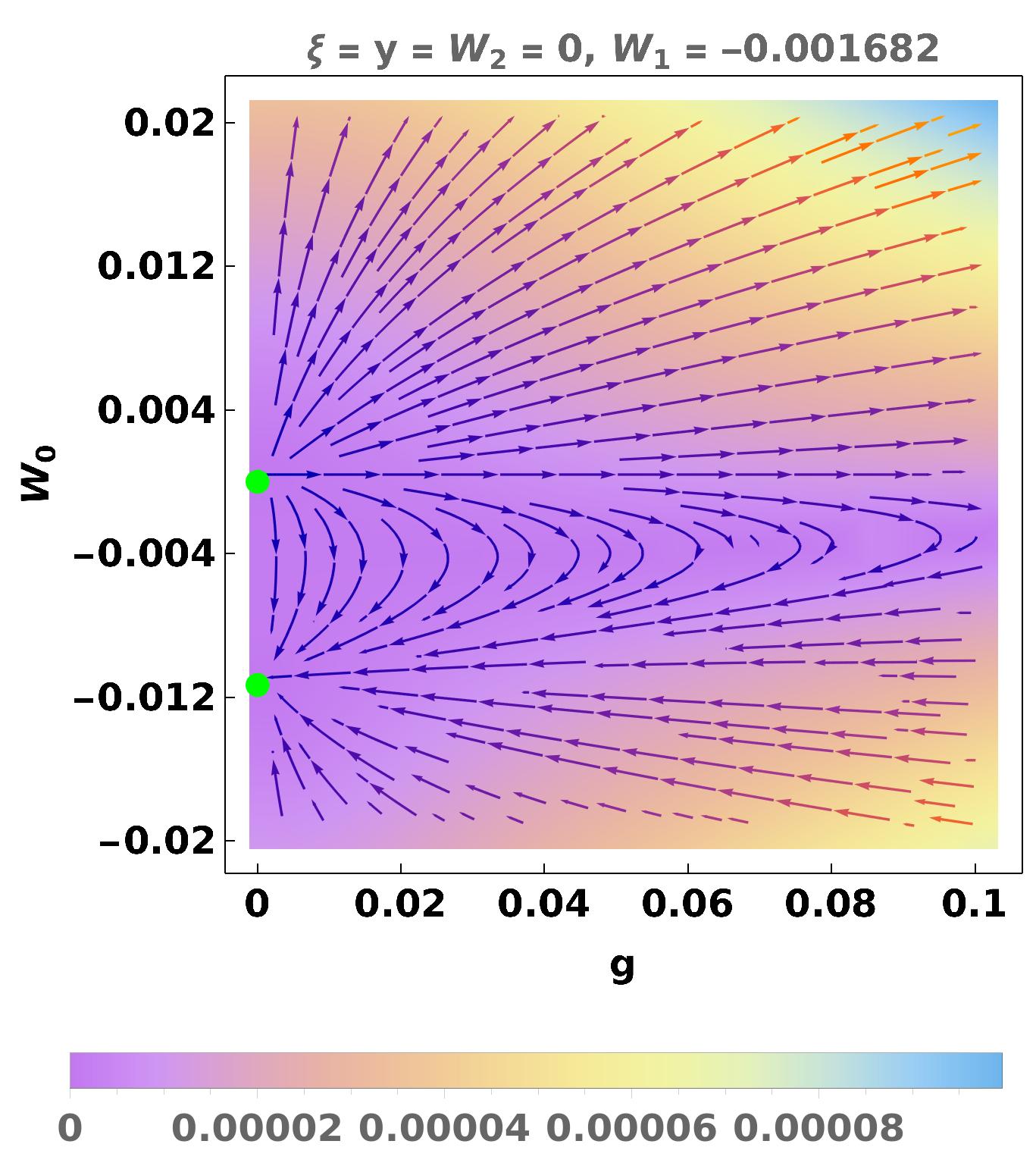}} \hspace{1 cm}
\subfigure[]{\includegraphics[width = 0.335 \columnwidth]{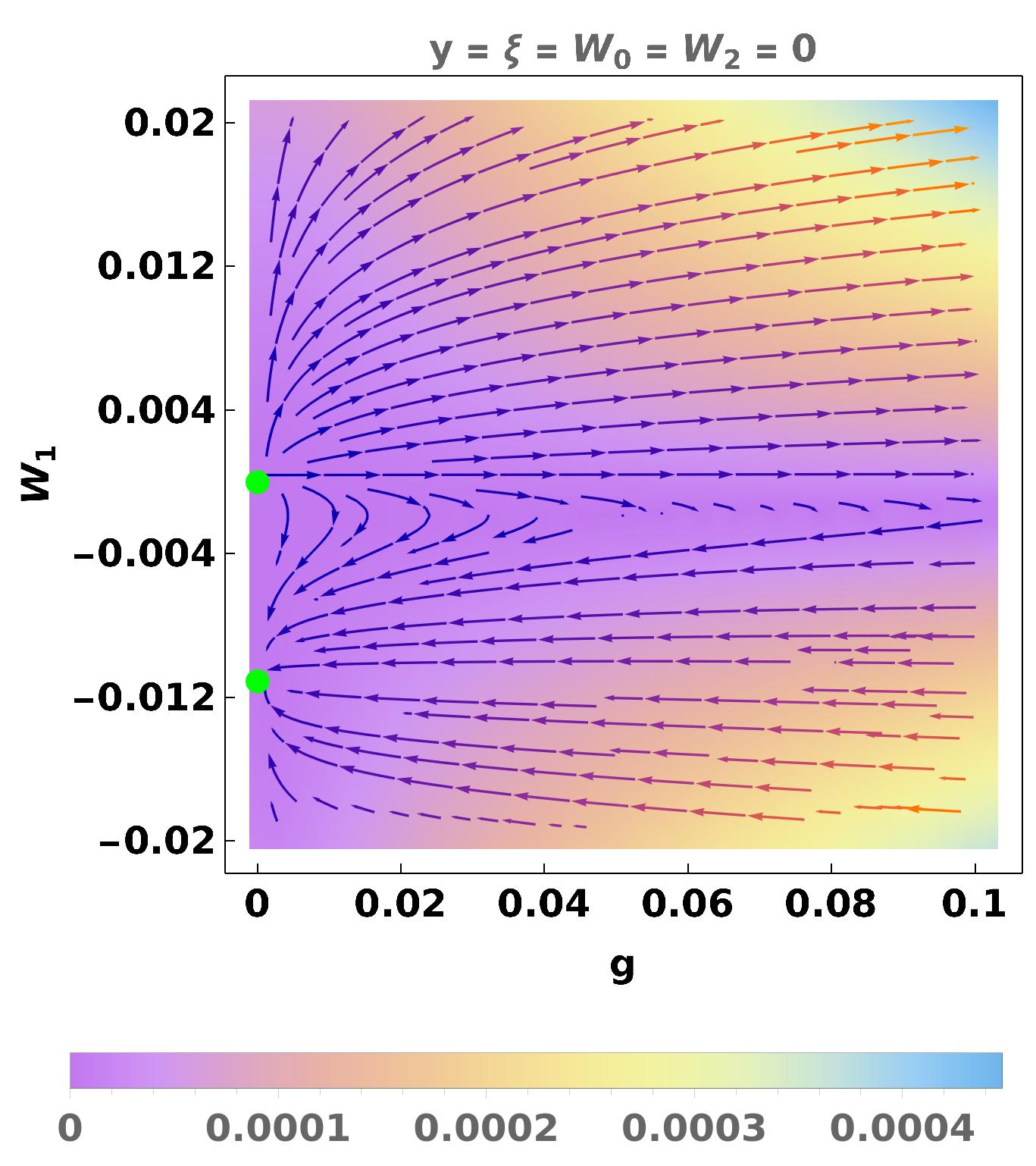}}
\caption{\label{disorder-fig}
The RG flow diagrams in the presence of disorder, for $N=4$ and
$\epsilon= 0.001$.
The green dots show the positions of the RG fixed points.}
\end{figure}
%%%%%%%%%%%%%%%%%%%%%%%%%%%

%%%%%%%%%%%%%%%%%%%%%%%%%%%%%%%%%%%
\subsection{Fixed points and their stability}

%%%%%%%%%%%%%%%%%%%%%%%%
Let us now examine the structure of the above RG equations in Eq.~\eqref{betafinal}. First, we note that
a fixed point with $\xi=y=0$ still exists, for which we can have either $g =0$ or $ |g|= 
\frac{12 \, \sqrt{6}\, \sqrt{3 \,W_0 +15 \,W_1 -10 \,W_2 +4 \,\pi ^2\, \epsilon }}
{\sqrt{ 107\,N-72}}$. For these fixed points, there are many allowed values of the $W_\alpha$'s.
Since the expressions for the fixed point values $\left (  |g^*| ,  W_0^*,  W_1^*, W_2^* \right )$ of the coupling constants for generic $N$ and $\epsilon$
are complicated, we list them for $N=4$ and $\epsilon=0.001$:
\begin{align}
& (0,	\,0,	\,0,	\,0),\quad
(0.309536,	\,0,	\,0,	\,0),\quad (0,\,	0,\,	-0.0109662,	\,0),\quad
(0.185944,	0,	-0.00168214,	0),\nn &
(0,\,	-0.0121472,\,	-0.0015184,	\,0),\quad
(0,	\,0.0131595,	\,-0.00657974,\,	0),
\quad (0,	0.00146036,	-0.00307,	-0.00145627),\nn &
(0,	\,0.00976256,\,-0.00399371,	\,-0.00101687),\quad
(0,\,	-0.0119048,	\,-0.00146232,\,	-0.000052146),\nn &
(0,	\,-0.000729334,	\,-0.00264258,\,	0.00164414),\quad
(0.111891,	\,0.000979658,	\,-0.00320206,\,	-0.0010772),\nn &
(0.124301,	\,0.00601502,	\,-0.00379805,	\,-0.000581355).
\end{align}

The stability of each fixed point can be determined from the linearized flow equations in the vicinity of the fixed point, which can be represented as:
\begin{align}
\frac{d}{d l}
\left(\begin{array}{c} 
\delta \xi\\ \delta y \\ \delta  g    \\ \delta W_0 \\ \delta W_1 \\ \delta W_2
\end{array}\right) \Bigg\rvert _{ \left ( \xi^*, y^*, g^* ,  W_0^*,  W_1^*, W_2^* \right ) }
 \approx 
\tilde{ \mathcal M}\left(\begin{array}{c} 
\delta \xi\\ \delta y \\ \delta  g  \\ \delta W_0 \\ \delta W_1 \\ \delta W_2
 \end{array}\right),
\end{align}
where $\tilde{ \mathcal M}$ is composed of the appropriate coefficients of the linearized equations.
Now, for $g=0$, $\beta_\xi $ is trivially zero, and a nonzero $\xi $ can be generated only via a nonzero $g$. Hence, for the $g=0$ case, there is always a zero eigenvalue for the $\xi$ direction.
%the stability analysis has to be performed by omitting the coupling constant $\xi$, in which case we have a $5\times 5$ stability matrix.
Evaluating the eigenvalues of the stability matrix, one can easily verify that
all the fixed points are unstable in the infrared, except the one with
$( \xi^*=0, \,y^*=0, \,g^* = 0,\,	W_0^* = 0,\,	W_1^*=-0.0109662,	\,W_2^*=0)$. 
The latter has an $\tilde{ \mathcal M}$ with five negative eigenvalues and one zero eigenvalue (corresponding to $\delta \xi$), implying marginal stability.

Some representative plots are shown in Fig.~\ref{disorder-fig}.
From the analysis of the RG flow equations, we find that for a negative initial value of $W_1$, there is a marginally stable fixed point at finite disorder couplings and zero $g$. However, for a positive initial value of $W_1$, all disorder couplings flow towards strong positive disorder, as the zero disorder fixed point is unstable. The nonzero $g$ fixed points are always unstable.
Hence, we conclude that the presence of disorder destroys the superconducting quantum critical point.

\section{Analysis and  discussion}
\label{discussion}

We have analyzed the effect of short-range correlated disorder on the superconducting quantum critical point in systems with quadratic band crossings in three dimensions. We have employed a perturbative RG framework in the minimal subtraction scheme. The problem includes all types of disorder as well as band-mass asymmetry (due to the term $\xi \, k^2$ in the Hamiltonian of Eq.~(\ref{bare})). We have found that disorder disrupts any possibility of getting a non-trivial stable superconducting quantum critical point at weak coupling. Furthermore, the system exhibits a runaway flow to strong disorder for any positive value of the vector disorder coupling.

The possibility of the conventional BCS type of superconductivity is already ruled out for these systems even in the clean limit, due to the vanishing density of states at the QBT point. Although superconductivity can occur at a finite coupling strength leading to a quantum critical point in the clean limit, presence of disorder completely destroys this as well.

We should remember that in our $\varepsilon$ expansion, we have to put $\varepsilon=1$ in the final results, in the same spirit as for the case of the Wilson-Fisher fixed point. We have not done higher loop calculations to check that the overall coefficients are significantly smaller than one-loop ones. Hence, our conclusions are suggestive taken into account the above fact. Furthermore, we neglected Coulomb interaction and it would be interesting to see the conclusions in presence of the Coulomb interaction. Lastly, in future works one can study the effect of cubic anisotropy in the scenario considered.

\section{Acknowledgments}
We thank Rahul M. Nandkishore, Shouvik Sur, Igor Boettcher, and Kush Saha for helpful discussions.

%%%%%%%%%%%%%%%%%%%%%%%%%%%%
\appendix

%%%%%%%%%%%%%%%%%%%%%
\section{Gamma matrix algebra}

 In this appendix, we list various identities which follow from the Clifford algebra. First, for $N_c$ gamma matrices ${\Gamma_a}$ $(a=1,2,\ldots, N_c )$, we have
\begin{align}
\sum \limits_{a} \Gamma_a\,\Gamma_a =N_c\,.
\label{rel0}
\end{align}

Other relations that have been used in various computations in the main text are:
%%%%%%%%%%%%%
\begin{align}
& \sum \limits_{a<b } \Gamma_{ab}\, \Gamma_{ab} =\frac{N_c \left(N_c -1\right )}{2} \,,\label{rel1}\\
&\Gamma_{f}\,\Gamma_{ab}\, \Gamma_{f} =\left (N_c -4\right ) \Gamma_{ab}\,,
\label{rel2}\\
%%%%%%%%%%%
& \Gamma_{cd} \,\Gamma_{ab} = \Gamma_{ab} \, \Gamma_{cd} 
+ 2 \,i \left( \delta _{ac}\, \Gamma_{db}  +  
\delta _{ad} \,\Gamma_{bc} + \delta _{bc}\,\Gamma_{ad} 
+ \delta_{bd} \, \Gamma_{ca} \right),
\label{rel3}\\
%%%%%%%%%%%%%%%%%%%%%%%%
& \sum \limits_a \Gamma_{ab}\, \Gamma_{ac}
= \left (N_c - 1 \right ) \Gamma_b \,\Gamma_c + i \,\Gamma_{cb}\,.
\label{rel4}
\end{align}

In our representation, where $\Gamma_{1,2,3}$ are real and $\Gamma_{4,5}$ are imaginary, we have:
\begin{align}
 \label{rg13} (\Gamma_a)^{\rm T} = \zeta_a \,\Gamma_a,\text{ where } \zeta_a =\begin{cases} 1 & a=1,2,3\\ -1 & a=4,5\end{cases},
\end{align}
which follows from $\Gamma_a^\dagger=\Gamma_a$.
Furthermore,
\begin{align}
 \label{rg14} \Gamma_{45}\, \Gamma_a = \zeta_a\, \Gamma_a\, \Gamma_{45},
\end{align}
following from the fact that $\Gamma_{45}$ is proportional to the product of $\Gamma_4$ and $\Gamma_5$.
Thus, we have
\begin{align}
 \label{rg15} 
 \Gamma _{45}^T=
 \frac{ \left( \Gamma _5 \right)^T \left(\Gamma _4\right)^T
 - \left(\Gamma _4\right)^T  \left( \Gamma _5\right)^T}  {2 \,i}
 =\frac{\Gamma _5 \,\Gamma _4-\Gamma _4 \,\Gamma _5}{2 \,i}
 =-\Gamma _{45}\,,\quad
 \Gamma_{45}\, (\Gamma_a)^{\rm T}   = \Gamma_a\, \Gamma_{45}  \,,
\end{align}
and
 \begin{align}
 \label{rg151} \Gamma_{45}\, (\Gamma_{a} \, \Gamma_{b} )^{\rm T}  
 = \Gamma_{45}\, \Gamma_b^{\rm T} \, \Gamma_a^{\rm T}  
 = \Gamma_b \, \Gamma_{45} \,\Gamma_a^{\rm T}
=  \Gamma_b  \,\Gamma_a \, \Gamma_{45}\,,\quad
%%%%%%%%%%%%
\Gamma_{45}\, (\Gamma_{ab} )^{\rm T}
= - \Gamma_{ab} \, \Gamma_{45}\,.
\end{align}

Since $N_c=5$ for the current problem, we can use the relations:
\begin{align}
& \Gamma_{a b}\, \Gamma_f =   
 i \left (  \delta_{a f} \, \Gamma_b - \delta_{b f} \,\Gamma_a \right ) - 
  \frac{ \epsilon_{abfcd}\, \Gamma_{cd} }{2} \,,
\quad
\Gamma_f \,\Gamma_{ab} = 
i \left (  \delta_{b f} \,\Gamma_a -\delta_{a f} \, \Gamma_b \right ) 
- \frac{ \epsilon_{abfcd}\, \Gamma_{cd} }{2}  \,,
%%%%%%%%%%%%%%%%%%%%
\\ &  \epsilon_{abcde}\,\epsilon_{abklm}
= 6 \left(  \delta_{dl}\,\delta_{em} -\delta_{dm} \,\delta_{el} \right),
%%%%%%%%%%%%%%%%%%%%%%%%%%
\\&  \epsilon_{abcde}\,\epsilon_{abclm} =
2 \left(    \delta _{cl}\, \delta _{ek}\, \delta _{dm}
+\delta _{ck}\, \delta _{dl} \, \delta _{em}
-\delta _{ck}\, \delta _{dm}\, \delta _{el}
-\delta _{cl}\, \delta _{dk}\, \delta _{em}
+\delta _{cm}\, \delta _{dk}\, \delta _{el}
-\delta _{cm}\, \delta _{dl}\, \delta _{\text{ek}} \right).
\label{rel41}
\end{align}
Using the above, we get:
\begin{align}
\label{rel51}
\sum \limits_{a <  b} \Gamma^i_{ab}\,  \Gamma^i_{e } 
\left (  \Gamma^j_{ab} \,  \Gamma^j_{e}  +  \Gamma^j_{e}  \,  \Gamma^j_{ab}\right )
& = 3 \sum \limits_{a< b} \Gamma_{ab}^i\, \Gamma_{ab}^j \,,
%%%%%%%%%%%%%%%%%%%%%%%%%
 \\ \label{rel52}\sum \limits_{a <  b,\,c, \, f} \Gamma^i_{ab}\,  \Gamma^i_{f} \,  \Gamma^i_{ c } 
\left ( \Gamma^j_{ab}  \,  \Gamma^j_{f} \,  \Gamma^j_{c }
+ \Gamma^j_{c } \,  \Gamma^j_{f} \,  \Gamma^j_{ab}  \right )
&= 34 \sum \limits_{a< b} \Gamma_{ab}^i\, \Gamma_{ab}^j \,,
%%%%%%%%%%%%%%%%%%%%
\\ \label{rel53}
\sum \limits_{a <  b,\,c<d, \, e} \Gamma^i_{ab}\,  \Gamma^i_{e} \,  \Gamma^i_{ cd } 
\left (  \Gamma^j_{ab } \,  \Gamma^j_{e} \,  \Gamma^j_{cd}  
+  \Gamma^j_{cd}  \,  \Gamma^j_{e} \,  \Gamma^j_{ab}\right )
& =  60 + 68 \sum \limits_{a} \Gamma_{a}^i\, \Gamma_{a}^j\,,
%%%%%%%%%%%%%%%%%%%%
\\ \label{rel54}
\sum \limits_{a <  b,\,c<d} \Gamma^i_{ab}\,   \Gamma^i_{ cd } 
\left (  \Gamma^j_{ab } \,    \Gamma^j_{cd}  
+  \Gamma^j_{cd}  \,   \Gamma^j_{ab}\right )
& =  26 +
12 \sum \limits_{a} \Gamma_{a}^i\, \Gamma_{a}^j
-12 \sum \limits_{a<b} \Gamma_{ab}^i\, \Gamma_{ab}^j\,.
\end{align}

%%%%%%%%%%%%%%%%%%%%%%%%%%%%%%%%%%%%%%%%%%%
\section{$d_a$-function algebra}
\label{angular}

We state some non-trivial relations for functions $d_a(\vec{p})$ derived in Ref.~\cite{igor16}.
Firstly, we have:
\begin{align}
 \sum_a d_a(\vec{p}) \, d_a(\vec{k}) = \frac{1}{d-1} \left [ d\times (\vec{p}\cdot\vec{k})^2 - p^2\, k^2 \right ].
\end{align}
For $\vec{k}=\vec{p}$ we obtain:
\begin{align}
 \label{d5} \sum_a d_a^2(\vec{p}) = p^4.
\end{align}
For $d=4$, we get:
\begin{align}
\label{d4} 
\sum_a  d_a(\vec{p})\, d_a(\vec{k}) =\frac{1}{3}\left [ 4\,(\vec{p}\cdot\vec{k})^2 - p^2\, k^2 \right  ] .
\end{align}

Due to reasons explained in Ref.~\cite{rahul-sid}, we have used the regularization scheme developed by Moon {\it et al} \cite{MoonXuKimBalents}. This involves continuing to four dimensions while keeping the angular and gamma matrix structure the same as in $d=3$. This translates into performing the radial momentum integrals with respect to a $d=4-\epsilon$ dimensional measure $\int \frac{p^{3-\epsilon} dp}{(2\pi)^{4-\epsilon}}$, but computing the angular momentum integrals only over the two-sphere parametrized by the polar and azimuthal angles ($\theta$ and $\varphi$). Nevertheless, the overall angular integral of an angle-independent function is taken to be $2\, \pi^2$ (since this is the total solid angle in  $d=4$), and hence, the angular integrals are normalized accordingly. Therefore, the angular integrations are performed with respect to the measure 
\be\label{moonmeasure}
\int dS \, \,(\ldots) \equiv \frac{\pi}{2} \int_0^{\pi} d \theta \int_0^{2\pi} d \varphi  \,   \sin \theta \,\,  (\ldots)\,,\ee
 where the $\pi/2$ is inserted for the sake of normalization. We refer to this as the ``Moon scheme".

Defining $d_{a}(\vec{p}) =  p^2\, \hat{d}_a(\vec{p})$, we have:
\begin{align}
& \int dS\, \hat{d}_{a}(\vec{p}) =0 \,,\label{dint}\\
& \int dS\, \hat{d}_{a}(\vec{p}) \,\hat{d}_{b}(\vec{p})  = \frac{ 2\, \pi^2 \, \delta_{ab}}{ N_c} \,.
\label{ddint}
\end{align}

%%%%%%%%%%%%%%%%%%%%%%%%%%%%%%%%%%%%%%%%%

\bibliography{disorder}
%============================================================================
%.............................................................................
%======================== END DOCUMENT +++++++++++++++++++++++++++++++++++++++
\end{document}